\documentclass[11pt,twoside,nohyper,preprint]{pallis}

\usepackage{bm}
\usepackage{epsfig}
\usepackage{graphics}

\usepackage{fancyh}
\usepackage{epsfig}
\usepackage{amssymb}
\usepackage{latexsym}
\usepackage{times}
\usepackage{slashed}
\usepackage{upgreek}
\usepackage{amsmath}
\usepackage{amsfonts}
\usepackage{amsbsy}
\usepackage{amscd}
\usepackage{bbm}
\usepackage{cite}

\newcommand{\eec}{\end{center}}
\newcommand{\bec}{\begin{center}}

\newcommand{\eem}{\end{matrix}}
\newcommand{\bem}{\begin{matrix}}
\newcommand{\eeq}{\end{equation}}
\newcommand{\beq}{\begin{equation}}
\newcommand{\ba}{\begin{array}}
\newcommand{\ea}{\end{array}}
\newcommand{\bea}{\begin{eqnarray}}
\newcommand{\eea}{\end{eqnarray}}
\newcommand{\baq}{\begin{eqnarray}}
\newcommand{\eaq}{\end{eqnarray}}
\newcommand{\beqs}{\begin{subequations}}
\newcommand{\eeqs}{\end{subequations}}
\newcommand{\bel}{\begin{align}}
\newcommand{\eal}{\end{align}}

\newcommand\eqs[2]{Eqs.~(\ref{#1}) and (\ref{#2})}

\newcommand{\ftn}{\footnotesize}

\newcommand{\TeV}{{\mbox{\rm TeV}}}
\newcommand{\MeV}{{\mbox{\rm MeV}}}
\newcommand{\GeV}{{\mbox{\rm GeV}}}
\newcommand{\eV}{{\mbox{\rm eV}}}
\newcommand{\EeV}{{\mbox{\rm EeV}}}
\newcommand{\PeV}{{\mbox{\rm PeV}}}
\newcommand{\ZeV}{{\mbox{\rm ZeV}}}
\newcommand{\YeV}{{\mbox{\rm YeV}}}
\newcommand{\ReV}{{\mbox{\rm ReV}}}

\newcommand{\hz}{{\mbox{\rm Hz}}}
\newcommand{\sFref}[2]{Fig.~\ref{#1}-{\small \sf ({#2})}}

\newcommand{\etal}{{\it et al.\/}}

\def\to{\rightarrow}

\def\llgm{\left\lgroup}
\def\rrgm{\right\rgroup}
\def\lf{\left(}
\def\rg{\right)}
\newcommand\vev[1]{\langle {#1} \rangle}
\newcommand\vevi[1]{\langle {#1} \rangle_{\rm I}}
\newcommand{\Gr}{\ensuremath{\widetilde{G}}}

\newcommand{\Nhi}{\ensuremath{N_{\rm I\star}}}
\newcommand{\ks}{\ensuremath{k_\star}}
\newcommand{\Gsm}{\ensuremath{\mathbb{G}_{\rm SM}}}

\newcommand{\Vhi}{\ensuremath{V_{\rm I}}}
\newcommand{\vf}{\ensuremath{V_{\rm F}}}
\newcommand{\Hhi}{\ensuremath{H_{\rm I}}}
\newcommand{\Whi}{\ensuremath{W_{\rm I}}}

\newcommand{\Vhio}{\ensuremath{V_{\rm I0}}}

\newcommand{\mP}{\ensuremath{m_{\rm P}}}

\newcommand{\Gbl}{\ensuremath{\mathbb{G}_{B-L}}}

\newcommand{\la}{\ensuremath{\lambda}}
\newcommand{\lm}{\ensuremath{\lambda_\mu}}

\newcommand{\aS}{\ensuremath{{\rm a}_S}}

\newcommand{\Gsn}{\ensuremath{\Gamma_{\rm I}}}
\newcommand{\msn}{\ensuremath{m_{\rm I}}}
\newcommand{\mgr}{\ensuremath{m_{3/2}}}
\newcommand{\mgri}{\ensuremath{m_{\rm I3/2}}}

\newcommand{\hd}{{\ensuremath{H_d}}}
\newcommand{\hu}{{\ensuremath{H_u}}}
\newcommand{\Nr}{\ensuremath{{\sf N}_{\rm G}}}

\newcommand{\ns}{\ensuremath{n_{\rm s}}}
\newcommand{\as}{\ensuremath{\alpha_{\rm s}}}
\newcommand{\sni}{\ensuremath{N^c_i}}

\newcommand{\bl}{\ensuremath{U(1)_{B-L}}}

\newcommand{\om}{\ensuremath{\omega}}
\newcommand{\ogw}{\ensuremath{\Omega_{\rm GW}h^2}}
\newcommand{\dN}{\ensuremath{\Delta N_{\rm eff}}}

\newcommand{\Dmax}{\ensuremath{\Delta_{\rm max*}}}
\newcommand{\Dex}{\ensuremath{\Delta_{\rm c\star}}}

\newcommand{\br}{\ensuremath{{\sf B}_{\rm h}}}

\newcommand{\Trh}{\ensuremath{T_{\rm rh}}}

\newcommand{\Tmax}{\ensuremath{T_{\rm max}}}
\newcommand{\sg}{\ensuremath{\sigma}}
\newcommand{\sgx}{\ensuremath{\sigma_\star}}
\newcommand{\sgc}{\ensuremath{\sigma_{\rm c}}}
\newcommand{\sgm}{\ensuremath{\sigma_{\rm max}}}

\newcommand{\ld}{\ensuremath{\lambda}}

\newcommand{\kp}{\ensuremath{\kappa}}

\def\ssb{\leavevmode\hbox{$\diagup$\kern-12pt\ftn\scshape
susy}}

\newcommand{\mss}{\ensuremath{\widetilde m}}

\newcommand{\Eref}[1]{Eq.~(\ref{#1})}
\newcommand{\Sref}[1]{Sec.~\ref{#1}}
\newcommand{\Fref}[1]{Fig.~\ref{#1}}
\newcommand{\Tref}[1]{Table~\ref{#1}}
\newcommand{\cref}[1]{Ref.~\cite{#1}}
\newcommand{\crefs}[1]{Refs.~\cite{#1}}

\def\th{{\theta}}
\def\ths{{\theta_S}}

\def\Ka{K\"{a}hler potential}
\def\Kaa{K\"{a}hler}
\def\Kam{K\"{a}hler manifold}
\def\Kap{K\"{a}hler potential}

\newcommand{\plk}{{\it Planck}}
\newcommand{\zm}{\ensuremath{Z_{-}}}

\newcommand{\bdhh}{{\ensuremath{\normalsize I{\kern-2.9pt H}}}}

\newcommand{\mD[1]}{\ensuremath{m_{#1\rm D}}}

\newcommand{\mM}{\ensuremath{m_{\rm M}}}

\newcommand{\Gm}{\ensuremath{\Gamma}}

\newcommand{\phc}{\ensuremath{\Phi}}
\newcommand{\phcb}{\ensuremath{\bar\Phi}}
\newcommand{\what}{\ensuremath{\widehat}}

\newcommand{\mgro}{\ensuremath{m_{3/2}}}
\newcommand{\mz}{\ensuremath{m_{z}}}
\newcommand{\mzi}{\ensuremath{m_{{\rm I}z}}}
\newcommand{\mth}{\ensuremath{m_{\theta}}}
\newcommand{\mthi}{\ensuremath{m_{\rm I\theta}}}

\newcommand{\no}{\ensuremath{N}}

\def\al{{\alpha}}
\def\bt{{\beta}}

\def\bz{{Z^*}}

\newcommand{\mcs}{\ensuremath{{G\mu_{\rm cs}}}}
\newcommand{\rcs}{\ensuremath{{r_{\rm cs}}}}
\newcommand{\ecs}{\ensuremath{{\epsilon_{\rm cs}}}}
\newcommand{\rms}{\ensuremath{{r_{\rm ms}}}}
\newcommand{\srms}{\ensuremath{{r_{\rm ms}^{1/2}}}}

\newcommand{\Ns}{\ensuremath{{N_{\rm I\star}}}}

\newcommand{\khh}{\ensuremath{K_{\rm H}}}

\newcommand{\dK}{\ensuremath{K_\mu}}

\newcommand{\dz}{\ensuremath{{\delta} z}}
\newcommand{\dzh}{\ensuremath{\what{\delta z}}}

\renewcommand{\Gsn}{\ensuremath{{\Gamma}_{\dz}}}
\newcommand{\Gth}{\ensuremath{{\Gamma}_{\th}}}
\newcommand{\Gh}{\ensuremath{{\Gamma}_{\tilde{h}}}}

\newcommand{\vtau}{\ensuremath{{\uptau}}}
\newcommand{\btau}{\ensuremath{\tilde\uptau}}
\newcommand{\btrh}{\ensuremath{\tilde\uptau_{\rm rh}}}
\newcommand{\btmax}{\ensuremath{\tilde\uptau_{\rm max}}}
\newcommand{\vteq}{\ensuremath{{\uptau_{\rm eq}}}}
\newcommand{\vtrh}{\ensuremath{{\uptau_{\rm rh}}}}
\newcommand{\vtmax}{\ensuremath{{\uptau_{\rm max}}}}
\newcommand{\vtmn}{\ensuremath{{\uptau_{\rm dc}}}}
\newcommand{\elg}{\ensuremath{{\ell_{\Gamma}}}}
\newcommand{\tk}{\ensuremath{{\text{k}}}}
\newcommand{\tz}{\ensuremath{{\text{z}}}}

\newcommand{\lrh[1]}{\ensuremath{\lambda_{#1N^c}}}

\def\nano{{\sf\ftn NANOGrav15}}
\def\ligo{{\sf\ftn LVK}}

\renewenvironment{subequations}{%
\refstepcounter{equation}%
\setcounter{parentequation}{\value{equation}}%
  \setcounter{equation}{0}
  \ignorespaces
}{%
  \setcounter{equation}{\value{parentequation}}%
  \ignorespacesafterend
}

\title{\LARGE\boldmath \bfseries\scshape PeV-Scale SUSY and Cosmic
Strings\\ from F-term Hybrid Inflation}

\author{\Large \bfseries\scshape C. Pallis\\
School of Civil Engineering, \\ Faculty of Engineering,\\
Aristotle University of Thessaloniki,  \\ GR-541 24 Thessaloniki,
GREECE\\ \vspace{3pt} \email{kpallis@auth.gr}}

\abstract{We consider F-term hybrid inflation (FHI) and SUSY
breaking in the context of a $B-L$ extension of MSSM which largely
respects a global $U(1)$ $R$ symmetry. The hidden sector Kaehler
manifold enjoys an enhanced $SU(1,1)/U(1)$ symmetry with the
scalar curvature determined by the achievement of a SUSY-breaking
de Sitter vacuum without ugly tuning. FHI turns out to be
consistent with data, provided that the magnitude of the emergent
soft tadpole term is confined in the range $(1.2-100)$ TeV, and it
is accompanied with the production of $B-L$ cosmic strings. If
these are metastable they interpret the present observations from
PTA experiments on the stochastic background of gravitational
waves with dimensionless tension $\mcs\simeq(1-9.2)\cdot10^{-8}$.
The $\mu$ parameter of MSSM arises by appropriately adapting the
Giudice-Masiero mechanism and facilitates the out-of-equilibrium
decay of the $R$ saxion at a reheat temperature lower than about
$71$ GeV. Due to the prolonged matter dominated era the
gravitational wave signal is suppressed at high frequencies. The
SUSY mass scale turns out to lie in the PeV region.

\\\\ \hspace*{10.5cm} {\it In memory of Prof. G.~Lazarides}

\\ \\ {\ftn\sffamily {\scshape Keywords}:  Cosmology, Inflation, Supersymmetric Models} \\
{\ftn\sffamily {\scshape PACS codes}:  98.80.Cq, 12.60.Jv,
95.30.Cq, 95.30.Sf}
\\\\ {\sl\bfseries Published in} {\sl Universe} {\bf 10}, no.~5, 211 (2024)}
\begin{document}

\setcounter{page}{1} \pagestyle{fancyplain}

\addtolength{\headheight}{.5cm}

\rhead[\fancyplain{}{ \bf \thepage}]{\fancyplain{}{\sc PeV-Scale
SUSY and CSs from FHI}} \lhead[\fancyplain{}{\sc
\leftmark}]{\fancyplain{}{\bf \thepage}} \cfoot{}

\section{Introduction}\label{intro}

\emph{Supersymmetric} ({\sf\ftn SUSY}) hybrid inflation
\cite{susyhybrid} based on F terms, called for short henceforth
\emph{F-term hybrid inflation} ({\sf\ftn FHI}) \cite{hinova} is
undoubtedly a well-motivated inflationary model -- for reviews see
\cref{lectures}. The most notable reasons which support our
statement above are the following:

\begin{itemize}

\item FHI is tied to a renormalizable superpotential uniquely
determined by a gauge and a global $U(1)$ $R$ symmetries;

\item FHI does not require fine-tuned parameters and
transplanckian inflaton values;

\item FHI can be reconciled with the \plk\ data \cite{plin} --
fitted to the standard power-law cosmological model with
\emph{Cold Dark Matter} ({\ftn\sf CDM}) and a cosmological
constant ($\Lambda$CDM) -- if we take properly into account not
only \emph{radiative corrections} ({\sf\ftn RCs}) but also
corrections originated by \emph{supergravity} ({\sf \ftn SUGRA})
\cite{pana, gpp, rlarge, kelar} as well as soft SUSY-breaking
terms \cite{mfhi, kaihi, sstad, split, ahmed24}.

\item FHI can be naturally followed by a \emph{Grand Unified
Theory} ({\sf \ftn GUT}) phase transition which may lead to the
production of cosmological defects, if these are predicted by the
symmetry breaking scheme. In the large majority of GUT breaking
chains the formation \cite{rachel} of \emph{cosmic strings}
({\sf\ftn CSs}) cannot be avoided.

\end{itemize}

Although the last feature above is often used to criticize the
powerfulness of the embedding of FHI in several GUTs -- see e.g.
\cref{univ} --, it recently appears as an interesting ingredient
in the cosmological model building. This is, because the announced
data from several \emph{pulsar timing array} ({\sf\ftn PTA})
experiments \cite{pta} -- most notably the \emph{NANOGrav 15-years
results} ({\sf\ftn \nano}) \cite{nano} -- provide strong support
for the discovery of a \emph{gravitational wave} ({\sf\ftn GW})
background around the nanohertz frequencies.

Given that the interpretation of this signal in terms of the
merging of supermassive black hole binaries is somewhat disfavored
\cite{nano1}, its attribution to gravitational radiation emitted
by topologically unstable superheavy CSs -- which may arise after
the end of FHI -- attracts a fair amount of attention \cite{nasri,
leont, so10, su5, buchfhi, infl}. In particular, the observations
can be interpreted if the CSs are meta- \cite{meta1} or
quasi-stable \cite{quasi}. Both types of topologically unstable
CSs arise from the symmetry breaking $\mathbb{G}\to \mathbb{H}
\times U(1)$, which produces monopoles. The subsequent $U(1)$
breaking yields CSs which connect monopoles with antimonopoles. We
here assume that these last ones are inflated away, but can appear
on metastable CSs via quantum pair creation. Therefore, metastable
CSs can be easily achieved if a $U(1)$ symmetry is embedded in a
gauge group with higher rank such as the Pati-Salam \cite{nasri},
the flipped $SU(5)$ \cite{leont} or $SO(10)$ \cite{so10}. For this
reason, in this work we focus on FHI realized in a $B-L$ extension
of \emph{Minimal SUSY Standard Model} ({\sf\ftn MSSM}) -- cf.
\cref{buchbl, nasri} -- which dilutes possibly preexisting
monopoles and is naturally accompanied by the production of a
network of CSs. On the other hand, we do not specify the mechanism
of the monopole production and the metastability of CSs, as done,
e.g., in \cref{buchfhi, infl}.


We adopt the minimal possible framework \cite{mfhi, kaihi} which
supports observationally acceptable FHI. It employs minimal \Ka\
for the inflaton field, RCs and soft SUSY-breaking terms. From the
last ones, the tadpole term plays a crucial role in establishing
compatibility with data. Its magnitude can be motivated by
intertwining \cite{asfhi} the \emph{inflationary sector} ({\sf
\ftn IS}) with a \emph{hidden sector} ({\sf \ftn HS}) introduced
in \cref{susyr}. Contrary to earlier attempts
\cite{buch1,nshi,high,stefan,davis} this HS respects a mildly
violated $R$ symmetry which is compatible with that adopted for
FHI \cite{susyhybrid}. The consequences of the interconnection of
the two sectors above are the following:

\begin{itemize}

\item The $R$ charge $2/\nu$ of the goldstino superfield -- which
is related to the geometry of the HS \cite{susyr} -- is
constrained to values with $0<\nu<1$.

\item The SUSY breaking is achieved not only in a Minkowski vacuum
-- as in the cases of \cref{buch1,nshi,high,stefan,davis} --  but
also in a \emph{de Sitter} ({\ftn\sf dS}) one which allows us to
control the notorious \emph{Dark Energy} ({\ftn\sf DE}) problem by
mildly tuning a single superpotential parameter to a value of
order $10^{-12}$.

\item The sgoldstino is stabilized \cite{buch1,high,stefan,davis}
to low values during FHI. This fact together with the selection of
a minimal \Ka\ for the inflaton assists us to resolve the
$\eta$-problem. Note that \Ka s inspired by the string theory are
mainly employed in earlier works
\cite{buch1,nshi,high,stefan,davis}.

\item The solution to the $\mu$ problem \cite{mubaer} of MSSM is
achieved by suitably applying \cite{susyr} the Giudice-Masiero
mechanism \cite{masiero, soft}. Contrary to similar attempts
\cite{dvali, muhi, nshi} the $\mu$ term here plays no role during
FHI but crucially controls the timely decay of the sgoldstino (or
$R$ saxion).

\item The energy density of the universe is dominated by the
energy density of sgoldstino condensate which decays \cite{baerh,
moduli, antrh, nsrh, full} before the onset of the \emph{Big Bang
Nucleosynthesis} ({\sf\ftn BBN}) at cosmic temperature
$(2-4)~\MeV$ \cite{nsref} thanks the aforementioned $\mu$ term.
Therefore, our scenario naturally motivates a prolonged matter
domination which causes a reduction \cite{wells, pillado} of the
spectrum of GWs at high frequencies ($f>0.1~{\rm Hz}$). This fact
is welcome since it assists us to avoid any conflict with the
third run advanced \emph{\small LIGO-VIRGO-KAGRA} ({\ftn\sf
\ligo}) data \cite{ligo}.


\item The SUSY mass scale $\mss$ is predicted to be close to the
$\PeV$ scale \cite{wellspev}. It fits well with the Higgs boson
mass, discovered at LHC, as it is estimated \cite{strumia} within
high-scale SUSY if we assume a relatively low $\tan\beta$ and stop
mixing. Note that the connection of inflation with SUSY breaking
has been extensively discussed in literature \cite{ant1} over the
last years.

\end{itemize}


In this feature paper we review further the model introduced in
\cref{asfhi} focusing exclusively on its implementation within a
version of MSSM endowed by a ${B-L}$ Higgs sector. We present the
complete particle content of the model, paying special attention
on the computation of the GWs emerging from the CS decay, under
the assumption that those are metastable. We also explain the
generation of neutrino masses taking into account SUGRA
contributions from \cref{tamv}.

The remaining manuscript is built up as follows: In \Sref{md} we
introduce our framework. Following, we revise the salient features
of our model as regards its vacuum in \Sref{des} and the
inflationary era in \Sref{fhi}. Then, we study the reheating
process in \Sref{rhs} and the production of the GWs from the CSs
in \Sref{css}. Our predictions for the SUSY-mass scale are exposed
in \Sref{susy}. Our conclusions are summarized in \Sref{con}.
Details for the derivation of neutrino masses are given in
Appendix~\ref{app}. Lastly, the acronyms adopted in the text are
arranged in Appendix~\ref{app1}.

\section{Model Set-up}\label{md}

We focus on an extension of MSSM invariant under the gauge group
$\Gbl= \Gsm\times U(1)_{B-L}$, where $\Gsm$ is the Standard Model
gauge group. The charge assignments under these symmetries of the
various matter and Higgs superfields are listed in
Table~\ref{tab1}. Namely, the $i$th generation $SU(2)_{\rm L}$
doublet left-handed quark and lepton superfields are denoted by
$Q_i$ and $L_i$ respectively, whereas the $SU(2)_{\rm L}$ singlet
antiquark [antilepton] superfields by $u^c_i$ and ${d_i}^c$
[$e^c_i$ and $\sni$] respectively. The electroweak Higgs
superfields which couple to the up [down] quark superfields are
denoted by $\hu$ [$\hd$]. Besides the MSSM particle content, the
model is augmented by seven superfields: a gauge singlet $S$,
three $\sni$'s, a pair of Higgs superfields $\phc$ and $\phcb$
which break $\bl$ and the goldstino superfield $Z$. In addition to
the local symmetry, the model possesses also the baryon and lepton
number symmetries and an $R$ symmetry $U(1)_{R}$. The latter plays
a crucial role to the construction of the superpotential (see
\Sref{md1}) and the \Kap\ (see \Sref{md2}).

\renewcommand{\arraystretch}{1.1}

\begin{table}[!t]
\begin{center}
\begin{tabular}{|c|c|c|c|c|}\hline
{\sc Superfields}&{\sc Representations}&\multicolumn{3}{|c|}{\sc
Global Symmetries}\\\cline{3-5}
&{\sc under $\Gbl$}& {\hspace*{0.3cm} $R$\hspace*{0.3cm} }
&{\hspace*{0.3cm}$B$\hspace*{0.3cm}}&{$L$} \\\hline\hline
\multicolumn{5}{|c|}{\sc Matter Superfields}\\\hline
{$e^c_i$} &{$({\bf 1, 1}, 1, 1)$}& $0$&$0$ & $-1$ \\
{$N^c_i$} &{$({\bf 1, 1}, 0, 1)$}& $0$ &$0$ & $-1$
 \\
{$L_i$} & {$({\bf 1, 2}, -1/2, -1)$} &$0$&{$0$}&{$1$}
\\
{$u^c_i$} &{$({\bf 3, 1}, -2/3, -1/3)$}& $0$  &$-1/3$& $0$
\\
{$d^c_i$} &{$({\bf 3, 1}, 1/3, -1/3)$}& $0$ &$-1/3$& $0$
 \\
{$Q_i$} & {$({\bf \bar 3, 2}, 1/6 ,1/3)$} &$0$ &$1/3$&{$0$}
\\ \hline
\multicolumn{5}{|c|}{\sc Higgs Superfields}\\\hline
{$\hd$}&$({\bf 1, 2}, -1/2, 0)$& {$2$}&{$0$}&{$0$}\\
{$\hu$} &{$({\bf 1, 2}, 1/2, 0)$}& {$2$} & {$0$}&{$0$}\\
\hline
{$S$} & {$({\bf 1, 1}, 0, 0)$}&$2$ &$0$&$0$  \\
{$\Phi$} &{$({\bf 1, 1}, 0, 2)$}&{$0$} & {$0$}&{$-2$}\\
{$\bar \Phi$}&$({\bf 1, 1}, 0, -2)$&{$0$}&{$0$}&{$2$}\\\hline
\multicolumn{5}{|c|}{\sc Goldstino Superfield}\\\hline
{$Z$}&$({\bf 1, 1}, 0,0)$&{$2/\nu$}&{$0$}&{$0$}\\
\hline\end{tabular}
\end{center}
\caption[]{\sl \small The representations under $\Gbl$ and the
extra global charges of the superfields of our model.}\label{tab1}
\end{table}
\renewcommand{\arraystretch}{1.}


\subsection{Superpotential}\label{md1}

The superpotential of our model respects totally the symmetries in
\Tref{tab1}. Most notably, it carries $R$ charge 2 and is linear
\emph{with respect to} ({\ftn\sf w.r.t.}) $S$ and $Z^{\nu}$. It
naturally splits into five parts:
\beq \label{Who} W=W_{\rm I} +W_{\rm H} +W_{\rm GH}+W_{\rm
MSSM}+W_{\rm MD},\eeq
where the subscripts ``I'' and ``H'' stand for the IS and HS
respectively and the content of each term is specified as follows:

\subparagraph{\sf\ftn (a)} $\Whi$ is the IS part of $W$ which
reads \cite{susyhybrid}
\beqs\beq \Whi = \kp S\left(\bar
\Phi\Phi-M^2\right),\label{whi}\eeq
where $\kp$ and $M$ are free parameters which may be made positive
by field redefinitions.

\subparagraph{\sf\ftn (b)} $W_{\rm H}$ is the HS part of $W$
written as \cite{susyr}
\beq W_{\rm H} = m\mP^2 (Z/\mP)^\nu. \label{wh} \eeq
Here $\mP=2.4~\ReV$ is the reduced Planck mass -- with
$\ReV=10^{18}~\GeV$ --, $m$ is a positive free parameter with mass
dimensions, and $\nu$ is an exponent which may, in principle,
acquire any real value if $W_{\rm H}$ is considered as an
effective superpotential valid close to the non-zero \emph{vacuum
expectation value} ({\ftn\sf v.e.v}) of $Z$, $\vev{Z}$. We assume
though that the effective superpotential is such that only
positive powers of $Z$ appear. If we also assume that $W$ is
holomorphic in $S$ then mixed terms of the form
$S^{\nu_s}Z^{\nu_z}$ can be forbidden in $W$ since the exponent of
a such term has to obey the relation
\bea
\nu_s+\nu_z/\nu=1~~\Rightarrow~~\nu_z=(1-\nu_s)\nu,\nonumber\eea
leading to negative values of $\nu_z$. This conclusion contradicts
with our assumptions above.


\subparagraph{\sf\ftn  (c)} $W_{\rm GH}$ is a term which mixes the
HS and the $B-L$ gauge fields of the IS. It has the form
\beq W_{\rm GH} = -\la\mP(Z/\mP)^\nu \phcb\phc \label{wgh} \eeq
with $\la$ a real coupling constant. The magnitude of $\la$ can be
restricted by the DE requirement as we see below.

\subparagraph{\sf\ftn (d)} $W_{\rm MSSM}$ is the part of $W$ which
contains the usual trilinear terms of MSSM, i.e.,
\beq W_{\rm MSSM} = h_{ijD} {d}^c_i {Q}_j \hd + h_{ijU} {u}^c_i
{Q}_j \hu+h_{ijE} {e}^c_i {L}_j \hd. \label{wmssm}\eeq
The selected $R$ assignments in \Tref{tab1} prohibit the presence
in $W_{\rm MSSM}$ of the bilinear $\mu\hu\hd$ term of MSSM and
other mixing terms -- e.g. $\lm S\hu\hd$ \cite{dvali} which is
frequently employed to generate this $\mu$ term. This term is
generated here via $\dK$ -- see \Sref{md2} below.

\subparagraph{\sf\ftn (e)} $W_{\rm MD}$ is the part of $W$ which
is provides masses to neutrinos
\beq W_{\rm MD} = h_{ijN} \sni L_j \hu+\lrh[i]\lf
S+(Z/\mP)^\nu\rg\phcb N^{c2}_i.\label{wmd}\eeq\eeqs
The first term in the right-hand side of \Eref{wmd} is responsible
for Dirac neutrino masses whereas for the Majorana masses -- cf.
\crefs{mfhi, univ}. The scale of the latter masses is intermediate
since $\vev{\phcb}\sim1~\YeV$ and $\vev{Z}\sim\mP$. The
cooperation of both terms lead to the light neutrino masses via
the well-known (type I) seesaw mechanism -- see also
Appendix~\ref{app}.

\subsection{K\"{a}hler Potential}\label{md2}

The \Ka\ respects the $\Gbl$, $B$ and $L$ symmetries in
\Tref{tab1}. It has the following contributions
\beq \label{Kho} K=K_{\rm I}+K_{\rm H}+\dK+K_{\rm
D}+|Y_\al|^2,\eeq
which left-handed chiral superfields of MSSM denoted by $Y_\al$
with $\al=1,...,7$, i.e.,
\bea Y_\al= {Q}, {L}, {d}^c, {u}^c, {e}^c, N^c,
\hd~\mbox{and}~\hu,\nonumber \eea
where the generation indices are suppressed.

\subparagraph{\sf\ftn (a)} $K_{\rm I}$ is the part of $K$ which
depends on the fields involved in FHI -- cf. \Eref{whi}. We adopt
the simplest possible choice  -- cf. \cref{mfhi, hinova} -- which
has the form
\beqs\beq K_{\rm I} = |S|^2+|\Phi|^2+|\bar\Phi|^2. \\
\label{ki} \eeq
Higher order terms of the form $|S|^{2\nu_S}/\mP^{2\nu_S-2}$ with
$\nu_S>1$ can not be excluded by the imposed symmetries but may
become harmless if $S\ll\mP$ and assume low enough coefficients.

\subparagraph{\sf\ftn (b)} $K_{\rm H}$ is the part of $K$ devoted
to the HS. We adopt the form introduced in \cref{susyr} where
\beq K_{\rm
H}=\no\mP^2\ln\lf1+\frac{|Z|^2-k^2\zm^4/\mP^2}{\no\mP^2}\rg
~~\mbox{with}~~Z_{\pm}=Z\pm Z^*. \label{khi} \eeq\eeqs
Here, $k>0$ mildly violates $R$ symmetry endowing $R$ axion with
phenomenologically acceptable mass. The selected $K_{\rm H}$ is
not motivated by the string theory but it can be considered as an
interesting phenomenological option for two reasons: It largely
respects the $R$ symmetry, which is a crucial ingredient for FHI,
and it ensures -- as we see in \Sref{des} -- a dS vacuum of the
whole field system with tunable cosmological constant for
\beq
\no=\frac{4\nu^2}{3-4\nu}~~\mbox{with}~~\frac34<\nu<\frac32~~\mbox{for}~~\no<0.\label{no}
\eeq
Our favored $\nu$ range finally is $3/4<\nu<1$. Since $\no<0$,
$K_{\rm H}$ parameterizes the $SU(1,1)/U(1)$ hyperbolic \Kam\ for
$k\sim0$.

\subparagraph{\sf\ftn (c)} $\dK$ includes higher order terms which
generate the needed mixing term between $\hu$ and $\hd$ in the
lagrangian of MSSM -- cf. \cref{masiero, susyr} -- and has the
form
\beqs\beq \dK=\lm\lf{\bz^{2\nu}}/{\mP^{2\nu}}\rg\hu\hd\ +\ {\rm
h.c.},\label{dK}\eeq
where the dimensionless constant $\lm$ is taken real for
simplicity.

\subparagraph{\sf\ftn (d)} $K_{\rm D}$ is an unavoidable term
which mixes the observable sector with the HS as follows
\beq K_{\rm D}=\ld_{ij\rm D}\lf Z^{*\nu}/\mP^{\nu+1}\rg\sni L_j
\hu+{\rm h.c}.\label{kd}\eeq\eeqs
It provides (subdominant) Dirac masses for $\nu_i$ \cite{tamv} as
shown in Appendix~\ref{app}.

\subparagraph{} The total $K$ in \Eref{Kho} enjoys an enhanced
symmetry for the $Y^\al, S$ and $Z$ fields, namely
\beq  \prod_\al U(1)_{Y^\al}\times U(1)_S\times  \lf
SU(1,1)/U(1)\rg_Z,\eeq
where the indices indicate the moduli which parameterize the
corresponding manifolds. Thanks to this symmetry, mixing terms of
the form $S^{\tilde \nu_s}Z^{*\tilde \nu_z}$ can be ignored
although they may be allowed by the $R$ symmetry for
$\tilde\nu_z=\nu\tilde\nu_s$. Most notably, $U(1)_S$ protects
$K_{\rm I}$ from $S$ depended terms which violates the $R$
symmetry, thereby, spoiling the inflationary set-up.

\section{\boldmath SUSY and $\Gbl$ Breaking -- Dark Energy}\label{des}

The vacuum of our model is determined by minimizing the F--term
(tree level) SUGRA scalar potential $V_{\rm F}$ derived
\cite{asfhi} from $W$ in Eq.~(\ref{Who}) and $K$ in \Eref{Kho}.
Note that D--term contributions to the total SUGRA scalar
potential vanish if we confine ourselves during FHI and at the
vacuum along the D-flat direction \beq
|\bar\Phi|=|\Phi|~~\mbox{which assures}~~ V_{\rm D}=
\frac{g^2}{2}\lf|\phc|^2-|\phcb|^2\rg^2=0. \label{inftr} \eeq Here
$g$ is the unique (considered unified) gauge coupling constant of
\Gbl and we employ the same symbol for the various superfields
$X^\al=S,Z,\Phi,\bar\Phi$ and their complex scalar components.

As we can verify numerically, $\vf$ is minimized at the
\Gbl-breaking vacuum
\beq \left|\vev{\Phi}\right|=\left|\vev{\bar\Phi}\right|=M.
\label{vevs} \eeq
It has also a stable valley along $\vev{\theta}=0$ and
$\vev{\ths/\mP}=\pi$, where these fields are defined by
\beq Z=(z+i\theta)/\sqrt{2}~~\mbox{and}~~S=\sg\
e^{i\ths/\mP}/\sqrt{2}.\label{Zpara}\eeq
Substituting \Eref{Zpara} in $\vf$ and minimizing it w.r.t the
various directions we arrive at the results
\beq \label{vevsg}\sg=-2^{(1-\nu)/2}\lf\ld(M^2+\mP^2)-m\mP\rg\
z^\nu/\mP^{(\nu+1)}~~\mbox{and}~~\vev{z}=2\sqrt{2/3}|\nu|\mP,\eeq
which yield the constant potential energy density
\beq \vev{\vf}=\lf\frac{16\nu^{4}}{9}\rg^\nu \lf\frac{\ld
M^2-m\mP}{\kp\mP^2}\rg^2\om^N\lf\ld(M^2+\mP^2)-m\mP\rg^2~~
\mbox{with}~~\om=e^{\frac{\vev{\khh}}{N\mP^2}}\simeq\frac{2(3-2\nu)}{3}.\label{vcc}\eeq
Tuning $\ld$ to a value $\ld\sim m/\mP\simeq10^{-12}$ we may wish
identify $\vev{\vf}$ with the DE energy density, i.e.,
\beq \label{omde} \vev{\vf}=\Omega_\Lambda\rho_{\rm
c}=7.3\cdot10^{-121}\mP^4,\eeq
where the density parameter of DE and the current critical energy
density of the universe are respectively given by \cite{plcp}
\beq \label{rhoc} \Omega_\Lambda=0.6889~~\mbox{and}~~\rho_{\rm
c}=2.31\cdot10^{-120}h^2\mP^4~~\mbox{with}~~h=0.6766.\eeq
Therefore, we obtain a post-inflationary dS vacuum which explains
the notorious DE problem. Moreover, \Eref{vevsg} yields
$\vev{\sg}\simeq0$.

The particle spectrum of the theory at the vacuum in
\eqs{vevs}{vevsg} includes the gravitino ($\Gr$) which acquires
mass \cite{susyr}
\beqs\beq \label{mgr} \mgro=\vev{e^{{\khh}/{2\mP^2}}W_{\rm
H}/\mP^2}\simeq 2^{\nu}3^{-\nu/2} |\nu|^{\nu}m\omega^{N/2}.\eeq
Diagonalizing the the mass-squared matrix of the field system
$S-\phc-\phcb-Z$ we also find out that the IS acquire a common
mass
\beq \msn=e^{{\khh}/{2\mP^2}}\sqrt{2}\lf\kp^2
M^2+(4\nu^{2}/3)^\nu(1+4M^2/\mP^2)m^2\rg^{1/2},\label{msn}\eeq
where the second term arises due to the coexistence of the IS with
the HS -- cf. \cref{mfhi}. Similar mixing does not appear in the
mass spectrum of HS which contains the (canonically normalized)
sgoldstino (or $R$ saxion) and the pseudo-sgoldstino (or $R$
axion) with respective masses
\beq \mz\simeq\frac{3\om}{2\nu}\mgro ~~\mbox{and}~~
\mth\simeq12k\om^{3/2}\mgro. \label{mzth}\eeq\eeqs

Applying finally the relevant formulas of \crefs{soft,susyr}, we
find that $\dK$ induces a non-vanishing $\mu$ term in the
superpotential of MSSM whereas $W_{\rm MSSM}$ and $\dK+|Y^\al|^2$
in \eqs{wmssm}{Kho} lead to a common soft SUSY-breaking mass
parameter $\mss$, at the vacuum of \Eref{vevsg}, which
indicatively represents the mass level of the SUSY partners.
Namely, we obtain
\beq W\ni \mu  H_u H_d ~~\mbox{with}~~|\mu|=
\lm\lf\frac{4\nu^2}{3}\rg^\nu(5-4\nu)\mgr~~\mbox{and}~~
\mss=\mgr.\label{mssi}\eeq
Variant form of the terms $|Y^\al|$ in $K$ -- see \Eref{Kho} -- do
not alter essentially our results \cite{susyr, asfhi}.

\section{Inflation Analysis}\label{fhi}

It is well known \cite{susyhybrid, lectures} that in global SUSY
FHI takes place for sufficiently large $|S|$ values along a F- and
D- flat direction of the SUSY potential
\begin{equation} \label{v0}\bar\Phi={\Phi}=0,~~\mbox{where}~~ V_{\rm SUSY}\lf{\Phi}=0\rg\equiv V_{\rm I0}=\kp^2
M^4~~\mbox{are}~~\Hhi=\sqrt{\Vhio/3\mP^2}\eeq
are the constant potential energy density and correspoding Hubble
parameter which drive FHI -- the subscript $0$ means that this is
the tree level value. In a SUGRA context, though, we first check
-- in Sec.~\ref{fhi1} -- the conditions under which such a scheme
can be achieved  and then in \Sref{fhi2} we give the final form of
the inflationary potential. Lastly, we present our results in
\Sref{fhi4} imposing a number of contraints listed in \Sref{fhi3}.

\subsection{Hidden Sector's Stabilization}\label{fhi1}

The attainment of FHI is possible if $Z$ is well stabilized during
it. The relative mechanism is pretty well known \cite{davis}. $Z$
is transported due to $V_{\rm I0}$ form its value in \Eref{vevs}
to a value well below $\mP$. To determine this, we construct the
complete expression for $\vf$ along the inflationary trajectory in
\Eref{v0} and then expand the resulting expression for low $S/\mP$
values, assuming that the $\theta=0$ direction is stable as in
\Eref{vevs}. Under these conditions $\vf$ is minimized for the
value
\beq \label{veviz}
\vevi{z}\simeq\lf\sqrt{3}\cdot2^{\nu/2-1}\Hhi/m\nu\sqrt{1-\nu}\rg^{1/(\nu-2)}\mP.\eeq
This result is in good agreement with its precise value derived
numerically. Note that $\nu<1$ assures a real value of $\vevi{z}$
with $\vevi{z}\ll\mP$ since $\Hhi/m\ll1$.

The (canonically normalized) components of sgoldstino, acquire
masses squared, respectively,
\beqs\beq\mzi^2\simeq6(2-\nu)\Hhi^2~~\mbox{and}~~
\mthi^2\simeq3\Hhi^2-
m^2\lf8\nu^2\mP^2-3\vevi{z}^2\rg\frac{4\nu(1-\nu)\mP^2+(1-96k^2\nu)\vevi{z}^2}{2^{3+\nu}\nu\mP^{2\nu}\vevi{z}^{2(2-\nu)}},
\label{mz8i} \eeq
whereas the mass of $\Gr$ turns out to be
\beq \mgri\simeq \lf
\nu(1-\nu)^{1/2}m^{2/\nu}/\sqrt{3}\Hhi\rg^{\nu/(2-\nu)}.\eeq\eeqs
It is clear from the results above that $\mzi\gg\Hhi$ and
therefore it is well stabilized during FHI whereas $m_{\rm
I\theta}\simeq\Hhi$ and gets slightly increased as $k$ increases.
However, the isocurvature perturbation is expected to be quite
suppressed since it becomes observationally dangerous only for
$\mthi\ll\Hhi$.

\subsection{Inflationary Potential}\label{fhi2}

Expanding $\vf$ for low $S$ values, introducing the canonically
normalized inflaton $\sg=\sqrt{2}|S|$ and taking into account the
RCs \cite{susyhybrid, lectures} we derive \cite{asfhi} the
inflationary potential $V_{\rm I}$ which can be cast in the form
\beq\label{vol} V_{\rm I}\simeq V_{\rm I0}\left(1+C_{\rm
RC}+C_{\rm SSB}+C_{\rm SUGRA}\right).\eeq
The individual contributions are specified as follows:

\subparagraph{\sf\ftn (a)} $C_{\rm RC}$ represents the RCs to
$V_{\rm I}/V_{\rm I0}$ which may be written as \cite{susyhybrid,
lectures,hinova}
\beqs\beq \label{crc}C_{\rm RC}=
{\kappa^2\over 128\pi^2}\lf8\ln{\kp^2 M^2\over
Q^2}+8x^2\tanh^{-1}\lf\frac{2}{x^2}\rg-4(\ln4-x^4\ln
x)+(4+x^4)\ln(x^4-4)\rg,\eeq
with
$x=(\sigma-\sqrt{2}\ld\vevi{Z}^\nu\mP^{1-\nu}/\kp)/M>\sqrt{2}$,
beyond which the expression above ceases to be valid. Here we take
into account that the multiplicity of the waterfall fields is
$\Nr=1$ since these are $U(1)_{B-L}$ non-singlets.

\subparagraph{\sf\ftn (b)} $C_{\rm SSB}$ is the contribution to
$V_{\rm I}/V_{\rm I0}$ from the soft SUSY-breaking effects
\cite{sstad} parameterized as follows:
\beq \label{cssb}C_{\rm SSB}=
m_{\rm I3/2}^2 \sg^2/2V_{\rm I0}-{\rm a}_S\,\sigma /\sqrt{2V_{\rm
I0}},\eeq
where the tadpole parameter reads
\beq \label{aSn} {\rm
a}_S=2^{1-\nu/2}m\frac{\vevi{z}^\nu}{\mP^\nu}\lf1+\frac{\vevi{z}^2}{2N\mP^2}\rg
\lf2-\nu-\frac{3\vevi{z}^2}{8\nu\mP^2}\rg.\eeq
The minus sign results from the stabilization of $\theta$ -- see
\Eref{Zpara} -- at zero and the minimization of the factor
$(S+S^*)=\sqrt{2}\sg\cos(\theta_S/\mP)$ which occurs for
$\theta_S/\mP=\pi~({\sf mod}~2\pi)$ -- the decomposition of $S$ is
shown in \Eref{Zpara}. We further assume that $\theta_S$ remains
constant during FHI so that the simple one-field slow-roll
approximation is valid. Possible variation of $\theta_S$ is
investigated in \cref{kaihi} where they found that acceptable
solutions with $\theta_S/\mP\neq\pi$ require a significant amount
of tuning. The first term in \Eref{cssb} does not play any
essential role in our set-up due to low enough $\mgr$'s --
cf.~\cref{mfhi}.

\subparagraph{\sf\ftn (c)} $C_{\rm SUGRA}$ is the SUGRA correction
to $V_{\rm I}/V_{\rm I0}$, after subtracting the one in $C_{\rm
SSB}$, which is
\beq \label{csugra} C_{\rm
SUGRA}=c_{2\nu}\frac{\sg^2}{2\mP^2}+c_{4\nu}\frac{\sg^4}{4\mP^4}~~\mbox{with}~~
 c_{2\nu}=\frac{\vevi{z}^2}{2\mP^2}~~\mbox{and}~~c_{4\nu}=\frac12\lf1+{\vevi{z}^2}{\mP^2}\rg.
\eeq\eeqs
Thanks to the minimality of $K_{\rm I}$ in \Eref{ki} and the
smallness of $\vevi{z}$ the coefficients above are low enough and
allow for FHI to be established -- cf.~\cref{sstad, mfhi}.

\subsection{Observational Requirements}\label{fhi3}

The analysis of FHI can be performed in the slow-roll
approximation if we calculate the slow-roll parameters \cite{plin}
\beq \label{sr} \epsilon={m^2_{\rm P}}\left(\frac{V'_{\rm
I}}{\sqrt{2}V_{\rm I}}\right)^2\simeq \frac{\mP^2}{2}\lf C'_{\rm
RC}+C'_{\rm SSB}\rg^2~~\mbox{and}~~\eta= m^2_{\rm
P}~\frac{V''_{\rm I}}{V_{\rm I}} \simeq\mP^2 C''_{\rm RC}, \eeq
where the derivatives of the various contributions read
\beqs\bel\label{cdev1} C'_{\rm SSB}&\simeq-\aS/\sqrt{2\Vhio},\\
\label{cdev2} C'_{\rm
RC}&\simeq\frac{kp^2x}{32M\pi^2}\lf4\tanh^{-1}\lf\frac{2}{x^2}
\rg+x^2\ln\lf1-\frac{4}{x^4}\rg\rg,\\
\label{cdev3} C''_{\rm
RC}&\simeq\frac{\kp^2}{32M^2\pi^2}\lf4\tanh^{-1}
\lf\frac{2}{x^2}\rg+3x^2\ln\lf1-\frac{4}{x^4}\rg\rg.
\end{align}\eeqs
The required behavior of $\Vhi$ in \Eref{vol} can be obtained
thanks to the relation $C'_{\rm RC}\simeq -C'_{\rm SSB}$  which is
established for carefully selecting $\kp$ (or $M$) and $\aS$.
Apparently, we have $C'_{\rm SSB}<0$ and $C'_{\rm RC}>0$ for
$\sgx<\sgm$ since
$|4\tanh^{-1}\lf{2}/{x^2}\rg|>|x^2\ln(1-4/x^4)|$. On the contrary,
$C''_{\rm RC}<0$, since the negative contribution
$3x^2\ln(1-4/x^4)$ dominates over the first positive one, and so
we obtain $\eta<0$ giving rise to acceptably low $\ns$ values.

Our model of FHI can be qualified if we test it against a number
of observational requirements. Namely:

\subparagraph{\sf\ftn (a)} The number of e-foldings elapsed
between the horizon crossing of the pivot scale $\ks=0.05/{\rm
Mpc}$ and the end of FHI has to be adequately large for the
resolution of the horizon and flatness problems of standard Big
Bang cosmology. Taking into account -- see \Sref{rhs} --, that FHI
is followed in turn, by a matter and a radiation dominated era,
the relevant condition takes the form \cite{hinova, plin}:
\begin{equation}  \label{Nhi}
\Ns=\int_{\sigma_{\rm f}}^{\sigma_{\star}} \frac{d\sigma}{m^2_{\rm
P}}\: \frac{V_{\rm I}}{V'_{\rm I}}\simeq19.4+{2\over
3}\ln{V^{1/4}_{\rm I0}\over{1~{\rm GeV}}}+ {1\over3}\ln {T_{\rm
rh}\over{1~{\rm GeV}}},
\end{equation}
where the prime denotes derivation w.r.t. $\sigma$, $\sgx$ is the
value of $\sigma$ when $\ks$ crosses outside the horizon of FHI
and $\sigma_{\rm f}$ is the value of $\sigma$ at the end of FHI.
This is normally obtained by the critical point
$\sgc=\sqrt{2}|S_{\rm c}|$ , i.e., the end of inflation coincides
with the onset of the $B-L$ phase transition. Note that
$\vev{\sg}\simeq0$, as mentioned below \Eref{omde}, and so it does
not disturb the inflationary dynamics which governs the $\sg$
evolution for $\sg\geq\sgc$.

\subparagraph{\sf\ftn (b)} The amplitude $A_{\rm s}$ of the power
spectrum of the curvature perturbation generated by $\sigma$
during FHI must be consistent with the data \cite{plcp} on
\emph{cosmic microwave background} ({\ftn\sf CMB}), i.e.,
\begin{equation} \label{Prob}
A_{\rm s}= \frac{1}{12\, \pi^2 m^6_{\rm P}}\; \left.\frac{V_{\rm
I}^{3}(\sigma_\star)}{|V'_{\rm I}(\sigma_\star)|^2}\right.\simeq\:
2.105\cdot 10^{-9}.
\end{equation}
The observed curvature perturbation is generated wholly by
$\sigma$ since the other scalars are massive enough during FHI --
see \Sref{fhi1}.

\subparagraph{\sf\ftn (c)} The remaining observables -- the scalar
spectral index $\ns$, its running $\as$, and the scalar-to-tensor
ratio $r$ -- which are calculated by the following standard
formulas
\beq \label{ns}  \ns=1-6\epsilon_\star\ +\ 2\eta_\star,
\as={2}\left(4\eta_\star^2-(\ns-1)^2\right)/3-2\xi_\star~~\mbox{and}~~
r=16\epsilon_\star,\eeq
(where $\xi\simeq m_{\rm P}^4~V'_{\rm I} V'''_{\rm I}/V^2_{\rm I}$
and all the variables with the subscript $\star$ are evaluated at
$\sigma=\sgx$) must be in agreement with data. We take into
account the latest Planck release 4 -- including TT,TE,EE+lowE
power spectra \cite{plcp} --, \emph{Baryon Acoustic Oscillations}
({\sf\ftn BAO}), CMB-lensing and BICEP/{\it Keck} data \cite{gws}.
Fitting it with $\Lambda$CDM$+r$ we obtain approximately
\beq \label{nswmap} \ns=0.965\pm0.009~~\mbox{and}~~r\lesssim0.032,
\eeq at 95$\%$ \emph{confidence level} ({\sf\ftn c.l.}) with
negligible $|\as|\ll0.01$.

\subsection{Results}\label{fhi4}

As deduced from \Sref{fhi1} -- \ref{fhi3}, the inflationary part
of our model depends on the parameters
\bea \kappa,~M,~m,~\ld,~\nu,~k~~\mbox{and}~~\lm.\nonumber \eea
Recall that $N$ is related to $\nu$ via \Eref{no}. Enforcing
\Eref{omde} fixes $\ld$  at a rather low value which does not
influence our remaining results. Moreover, $k$ affects exclusively
$\mth$ and $\mthi$ via \eqs{mzth}{mz8i}. We select throughout the
value $k=0.1$ which assures the avoidance of massless modes. We
here present the $\aS$ values as a function of $\kp$ or $M$ which
assist the confrontation of FHI with data -- cf. \cref{mfhi} --
and postpone their derivation from $m$ and $\nu$ via \Eref{aSn} in
\Sref{susy}. As regards $\Trh$, which influences \Eref{Nhi}, we
adopt a value close to those met in our set-up, $T_{\rm
rh}\simeq1~\GeV$.

\renewcommand{\arraystretch}{1.25}
\begin{table}[!t] \bec
\begin{tabular}{|c|c|c||c|c|c|}\hline
{\sc Benchmark}& & &{\sc Benchmark} & & \\
{\sc Point:}& A & B&{\sc Point:} &A &B \\\hline\hline
\multicolumn{6}{|c|}{\sc Inputs}\\
\hline
$M~(\YeV)$&$1.4$&$2.1$&$\kp$&$0.0005$&$0.001$\\
$m~(\PeV)$&$0.5$&$3.5$&$\ld~(10^{-12})$&$0.2$&$1.4$\\ \hline
\multicolumn{6}{|c|}{\sc Inflationary Parameters}\\\hline
$\aS~(\TeV)$&$2.63$&$25.3$&$\Dex~(\%)$&$2.6$&$8.2$\\
$\Hhi~(\EeV)$&$0.23$&$1.05$&$\sgm/M$&$1.49$&$1.66$\\
$\sgx/M$&$1.45$&$1.53$&$\Dmax~(\%)$&$2.7$&$7.7$\\
$\Ns$&$40.5$&$40.8$&$\sg_{\rm min}/M$&$35.5$&$30.8$\\\hline
\multicolumn{6}{|c|}{\sc Observables}\\\hline
$\ns$&\multicolumn{2}{c||}{$0.965$}&$-\as~(10^{-4})$&$2.4$&$3.1$\\\cline{2-3}
$r$&$9\cdot10^{-13}$&$1.8\cdot10^{-11}$&$\mcs~(10^{-7})$&$2.3$&$6$\\\hline
\multicolumn{6}{|c|}{\sc $z$ v.e.v and Particle Spectrum}\\\hline
\multicolumn{3}{|c||}{\sc During FHI}&\multicolumn{3}{c|}{\sc at
the Vacuum $\vev{z}=1.4\mP$}\\\hline
$\vevi{z}~(10^{-3}\mP)$&$1.3$&$2$&$m_{\rm 3/2}~(\PeV)$&$0.9$&$6.2$\\
$m_{\rm I3/2}~(\TeV)$&$1.2$&$11.2$&$m_{z}~(\PeV)$&$1.2$&$8.8$\\
$m_{{\rm I}z}~(\EeV)$&0.6&$2.7$&$m_{\theta}~(\PeV)$&$0.9$&$5.6$\\
$m_{\rm I\theta}~(\EeV)$&$0.08$&$0.5$&$m_{\rm
I}~(\ZeV)$&$1.7$&$5.2$\\\hline
\multicolumn{6}{|c|}{\sc Reheating Process}\\\hline
$\mu/\mss$&\multicolumn{2}{|c||}{\sc $T_{\rm
max}~(\PeV)$}&$\mu/\mss$&\multicolumn{2}{c|}{\sc
$-\vtmax$}\\\hline
$3$&$0.3$&$2.2$&$3$&$66.6$&$67$\\
$1/3$&$0.1$&$0.7$&$1/3$&$67.4$&$67.6$\\\hline
$\mu/\mss$&\multicolumn{2}{|c||}{\sc $\Trh
(\GeV)$}&$\mu/\mss$&\multicolumn{2}{c|}{$-\vtrh$}\\\hline
$3$&$0.21$&$3.5$&$3$&$28.3$&$31.3$\\
$1/3$&$0.04$&$0.43$&$1/3$&$26.3$&$29.1$\\\hline
\end{tabular}
\end{center}
\caption[]{\sl\small A benchmark table of our scenario. We fix
$\nu=7/8$ (resulting to $\no=-49/8$) and $k=0.1$. Recall that
$1~\PeV=10^6~\GeV$, $1~\EeV=10^9~\GeV$, $1~\ZeV=10^{12}~\GeV$ and
$1~\YeV=10^{15}~\GeV$.} \label{tab}
\end{table}\renewcommand{\arraystretch}{1.}

Enforcing  Eqs.~(\ref{Nhi}) and (\ref{Prob}) we can restrict $M$
and $\sgx$ as functions of our free parameters $\kappa$ and $\aS$.
It is apparent that the allowed ranges of parameters is similar to
those explored in \cref{mfhi}, where the HS is not specified. Some
deviations are only due to the improvements on the determination
of the $\ns$ values in \Eref{nswmap}. The correct values of that
quantity are attained if FHI becomes of hilltop type \cite{mfhi,
kaihi} . I.e., $\Vhi$ is non-monotonic and develops a maximum at
$\sgm$ and a minimum at $\sg_{\rm min}\gg\sgm$. For $\sg>\sgm$
$\Vhi$ becomes a monotonically increasing function of $\sg$ and so
its boundedness is assured. FHI takes place for $\sg<\sgm$. The
position of $\sgm$ is predominantly dominated by $C'_{\rm RC}$ and
$C'_{\rm SSB}$. On the other hand, $\sg_{\rm min}$ can be roughly
found by the interplay of $C'_{\rm SUGRA}$ and $C'_{\rm SSB}$.
Namely, we obtain \cite{asfhi}
\begin{equation}
\label{sgm} \sgm\simeq
\frac{\kp^3M^2}{4\sqrt{2}\pi^2\aS}~~\mbox{and}~~\sigma_{\rm
min}\simeq \lf\frac{\aS\mP^4}{\sqrt{2}c_{4\nu}\kp
M^2}\rg^{1/3}\cdot\eeq
Note that $\sgm$ is independent from $C_{\rm SUGRA}$ and
$\sigma_{\rm min}$ from $C_{\rm RC}$. The attainment of successful
hilltop FHI requires the establishment of the hierarchy
$\sgc<\sgx<\sgm$. From our numerical computation we observe that,
for constant $\kp$ and $\aS$, $n_{\rm s}$ decreases as $\sgx$
approaches $\sgm$. To quantify somehow the amount of these
tunings, we define the quantities
\beq \Dex={\sgx/\sgc}-1~~\mbox{and}~~\Dmax=
{1-\sgx/\sgm}\,.\label{dms}\eeq
The naturalness of the hilltop FHI increases with $\Dex$ and
$\Dmax$.

To get an impression of the required values of the parameters of
the model, we construct first the benchmark table \Tref{tab}. We
display there the results of our analysis for two \emph{benchmark
points} ({\ftn\sf BPs}) with $\kp=0.0005$ (BPA) and $\kp=0.001$
(BPB) and $\ns$ fixed to its central value in \Eref{nswmap}. We
also employ some representative $\nu$ and $k$ values. In all
cases, we obtain $\Nhi\sim40$ from \Eref{Nhi}. From the
observables listed in \Tref{tab} we infer that $|\as|$ turns out
to be of order $10^{-4}$, whereas $r$ is extremely tiny, of order
$10^{-11}$, and therefore far outside the reach of the forthcoming
experiments devoted to detect primordial GWs. From the entries of
\Tref{tab} related to $\Dex$ and $\Dmax$, we notice that
$\Dmax>\Dex$, their values may be up to $10\%$, and increase as
$\aS$ or $\kp$ (and $M$) increases. From  the mass spectra
arranged in \Tref{tab} -- see \Sref{fhi1} -- we see that $\mgri$
is similar to $\aS\sim\TeV$ whereas the other masses are of order
$\EeV$ whereas at the vacuum $\msn$ turns out to be of order
$1~\ZeV=10^{12}~\GeV$ whereas $\mgr, \mz$ and $\mth$ lie in the
$\PeV$ range -- see \Sref{des}. In the same Table we find it
convenient to accumulate the values of some parameters related to
the reheating process, specified in Sec.~\ref{rhs} below, and the
formed CSs -- see \Sref{css}.

\begin{figure}[!t]\vspace*{-.12in}
\hspace*{-.19in}
\begin{minipage}{8in}
\epsfig{file=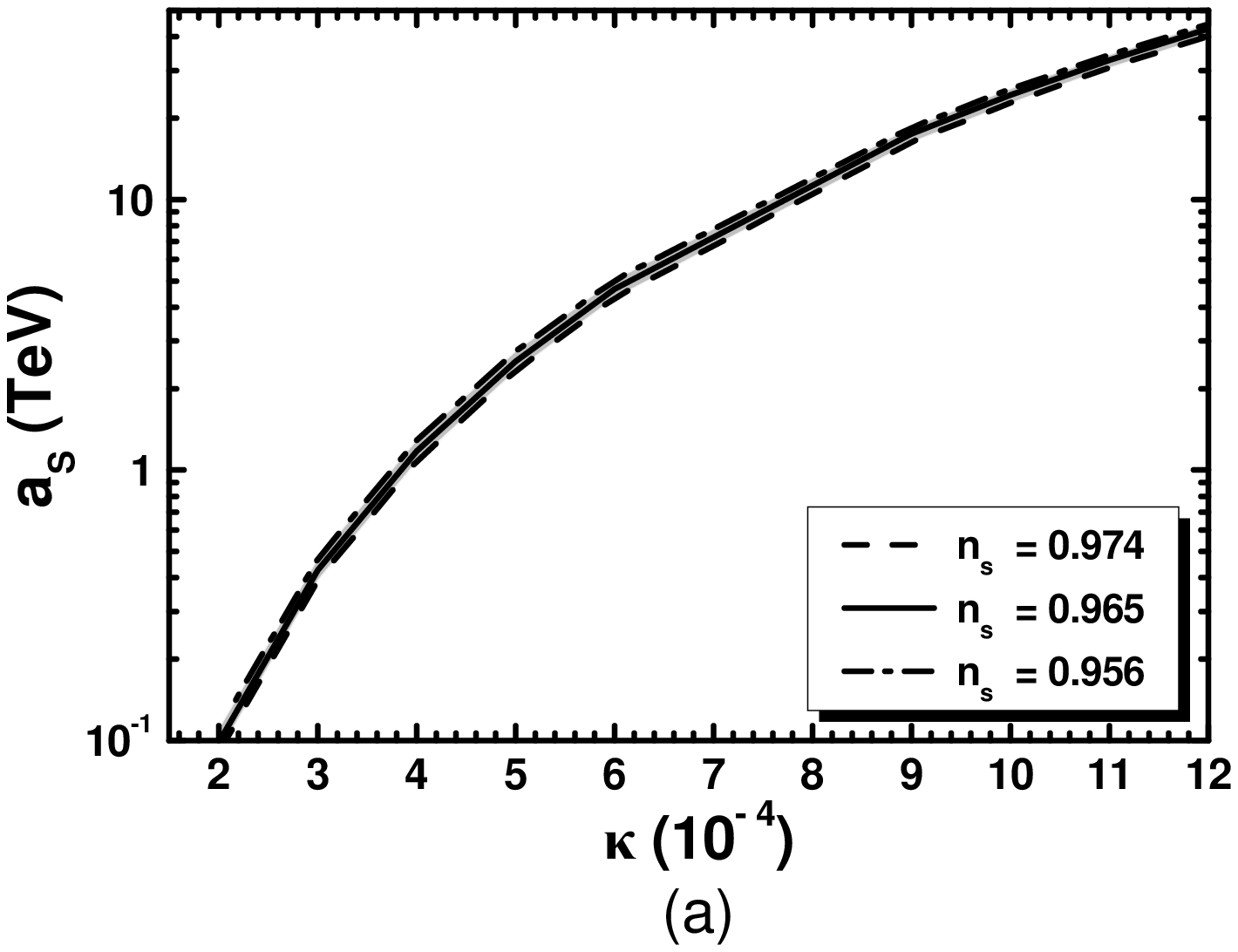,height=3.6in,angle=-90}
\hspace*{-1.2cm}
\epsfig{file=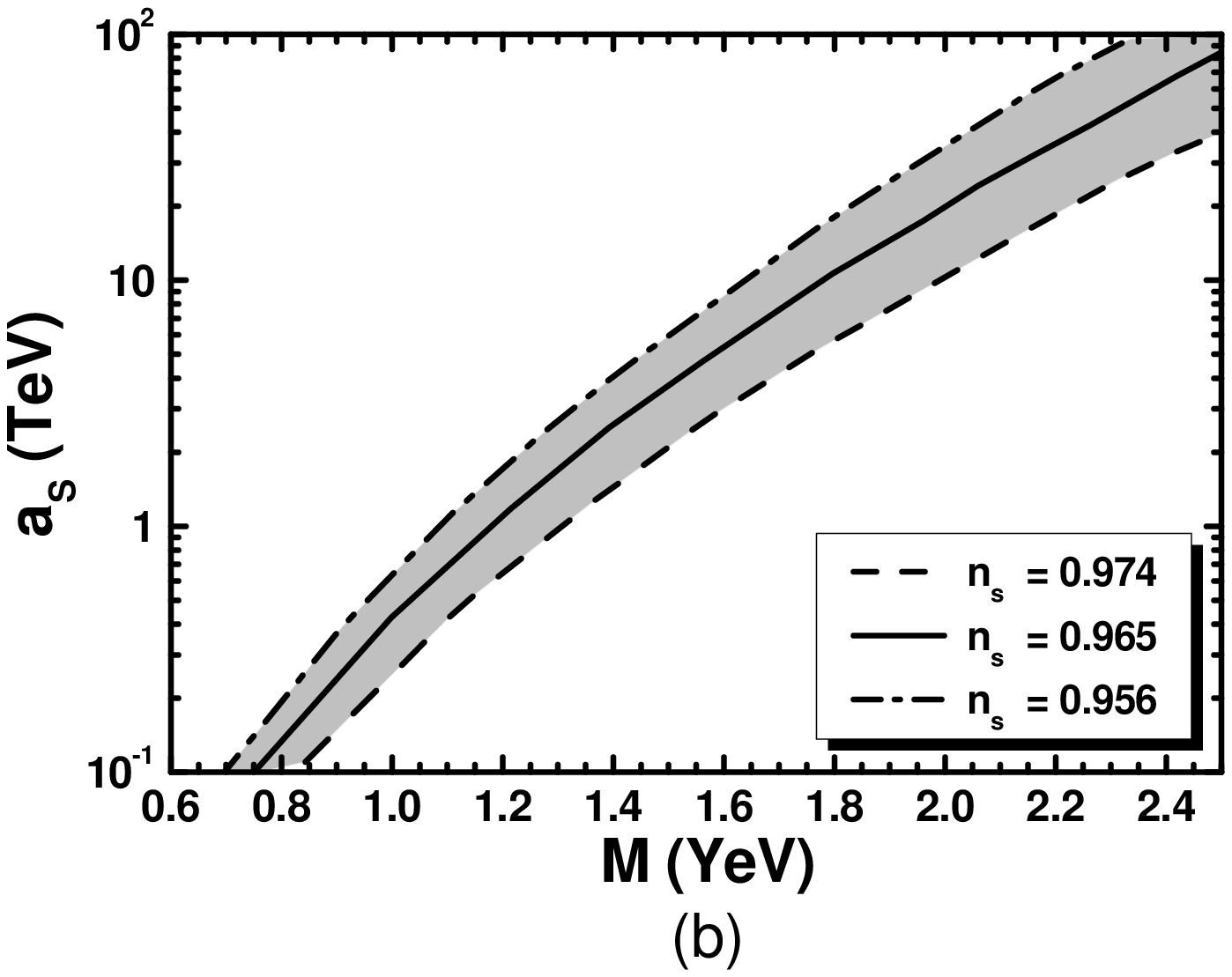,height=3.6in,angle=-90} \hfill
\end{minipage}
\hfill \caption{\sl\small Allowed (shaded) regions as determined
by Eqs.~(\ref{Nhi}), (\ref{Prob}), and (\ref{nswmap}) in the
$\kp-\aS$ {\sffamily\ftn (a)} and $M-\aS$ {\sffamily\ftn (b)}
plane. The conventions adopted for the various lines are also
shown. }\label{fig2}
\end{figure}

To explore further the parameter space of our model allowed by the
inflationary requirements in Eqs.~(\ref{Nhi}), (\ref{Prob}) and
(\ref{nswmap}) we present the gray shaded regions in the
$\kappa-\aS$ and $M-\aS$ plane  -- see \sFref{fig2}{a} and
\sFref{fig2}{b}.  The boundaries of the allowed areas in
\Fref{fig2} are determined by the dashed [dot-dashed] lines
corresponding to the upper [lower] bound on $n_{\rm s}$ in
Eq.~(\ref{nswmap}) -- note that in \sFref{fig2}{a} we obtain a
very narrow strip from the $n_{\rm s}$ variation, We also display
by solid lines the allowed contours for $\ns=0.965$. The maximal
$r$'s are encountered in the upper right end of the dashed lines
-- corresponding to $\ns=0.974$. On the other hand, the maximal
$|\as|$'s are achieved along the dot-dashed lines and the minimal
value is $\as=-3.2\cdot10^{-4}$. Summarizing our findings from
Fig.~\ref{fig2} for central $n_{\rm s}$ in \Eref{nswmap} we end up
with the following ranges:
\beq0.07\lesssim {M/{\rm
YeV}}\lesssim2.6~~\mbox{and}~~0.1\lesssim{\aS/\TeV}\lesssim100.
\label{res1} \eeq
Within the margins above, $\Dex$ ranges between $0.5\%$ and
$9.5\%$ and $\Dmax$ between $0.4\%$ and $8.2\%$ with the relevant
values increasing with $M$ and $\aS$, as deduced from \Tref{tab}
too. The lower bounds of the inequalities above are expected to be
displaced to slightly upper values due to the post-inflationary
requirements -- see \Eref{tns} below -- which are not considered
here for the shake of generality.

\section{Reheating Process}\label{rhs}

Soon after FHI, the IS and $z$ enter into an oscillatory phase
about their minima in \eqs{vevs}{vevsg} and eventually decay via
their coupling to lighter degrees of freedom. Note that $\theta$
and $\ths$ remain well stabilized at their values shown below
\Eref{vevs} during and after FHI and so they do not participate in
the phase of damped oscillations. The commencement of
$z$-dominated phase occurs for values of Hubble rate $H_{z\rm
I}\sim\mz$. Given that $\vev{z}\sim\mP$ -- see \Eref{vevsg} --,
the initial energy density of its oscillations $\rho_{z\rm I}$ is
comparable with the energy density of the universe $\rho_{z\rm
It}$ at the onset of these oscillations since
\beq \rho_{z\rm I}\sim\mz^2\vev{z}^2~~\mbox{and}~~\rho_{z\rm
It}=3\mP^2H_{z\rm I}^2\simeq3\mP^2\mz^2.\eeq
Therefore, we expect that $z$ will dominate the energy density of
the universe until completing its decay through its weak
gravitational interactions. Due to weakness of these interactions
we expect that the reheating temperature $\Trh$ will be rather
low. This is the notorious cosmic moduli problem
\cite{baerh,moduli} which plagues the vast majority of the SUGRA
settings.

In our model though $\Trh$ adequately increases thanks to the
large enough $\mz$ and $\mu$ originated by $\dK$ in \Eref{dK}. To
show this fact we estimate $\Trh$ by \cite{rh}
\beq \label{Trh} \Trh= \left({72/5\pi^2g_{\rm
rh*}}\right)^{1/4}\Gsn^{1/2}\mP^{1/2},\eeq where $g_{\rm
rh*}\simeq10.75-100$ counts for the effective number of the
relativistic degrees of freedom at $\Trh$. This is achieved for
\beq \btrh\simeq-
\frac{2}{3}\ln\frac{2}{5}\sqrt{3}\mP\Gsn\rho_{z\rm
I}^{-1/2},\label{btrh} \eeq
where $\btau=\ln\left(R/R_{\rm I}\right)$ with $R$ the scale
factor of the universe and $R_{\rm I}$ its value at the onset of
the $z$ oscillations. Also $\Gsn$ is the total decay width of the
(canonically normalized) sgoldstino
\beq\dzh=\vev{K^{1/2}_{ZZ^*}}\dz~~\mbox{with}~~\dz=z-\vev{z}~~\mbox{and}~~\vev{K_{ZZ^*}}=\vev{\om}^{-2}
\label{dphi}\eeq
which predominantly includes the contributions from its decay into
pseudo-sgoldstinos, $\theta$ and electroweak higgs, $\hu$ and
$\hd$ via the kinetic terms $K_{XX^*}\partial_\mu X\partial^\mu
X^*$ with $X=Z,\hu$ and $\hd$ \cite{full,baerh,antrh,nsrh} of the
Lagrangian. In particular, we have
\beq\Gsn\simeq\Gth+\Gh,\label{Gol}\eeq
where the individual decay widths are found to be
\beq \Gth\simeq\frac{\ld_\theta^2\mz^3}{32\pi
\mP^2}\sqrt{1-\frac{4\mth^2}{\mgr^2}}~~\mbox{with}~~\ld_\theta=\frac{\vev{z}}{N}=\frac{4\nu-3}{\sqrt{6}\nu}
~~\mbox{and}~~\Gh=\frac{2^{4\nu-1}}{3^{2\nu-1}}\lm^2\frac{\om^2}{4\pi}
\frac{\mz^3}{\mP^2}\nu^{4\nu}\,.\label{Gth}\eeq
From the expressions above we readily recognize that $\Gsn$ is
roughly proportional to $\mz^3/\mP^2$ as expected for any typical
modulus \cite{baerh,moduli}. We explicitly checked that $z$-decay
channels into gauge bosons through anomalies and three-body MSSM
(s)particles are subdominant. On the other hand, we kinematically
block the decay of $\dzh$ into $\Gr$'s selecting $\nu>3/4$ which
ensures $\mz<2\mgr$ -- see \Eref{mzth}. We do so in order to
protect our setting from the so-called \cite{koichi}
moduli-induced $\Gr$ problem, i.e., the possible late decay of the
produced $\Gr$ and the problems with the abundance of the
subsequently produced lightest SUSY particles -- cf. \cref{baerh}.

\begin{figure}[!t]\vspace*{-.25in}
\bec\includegraphics[width=60mm,angle=-90]{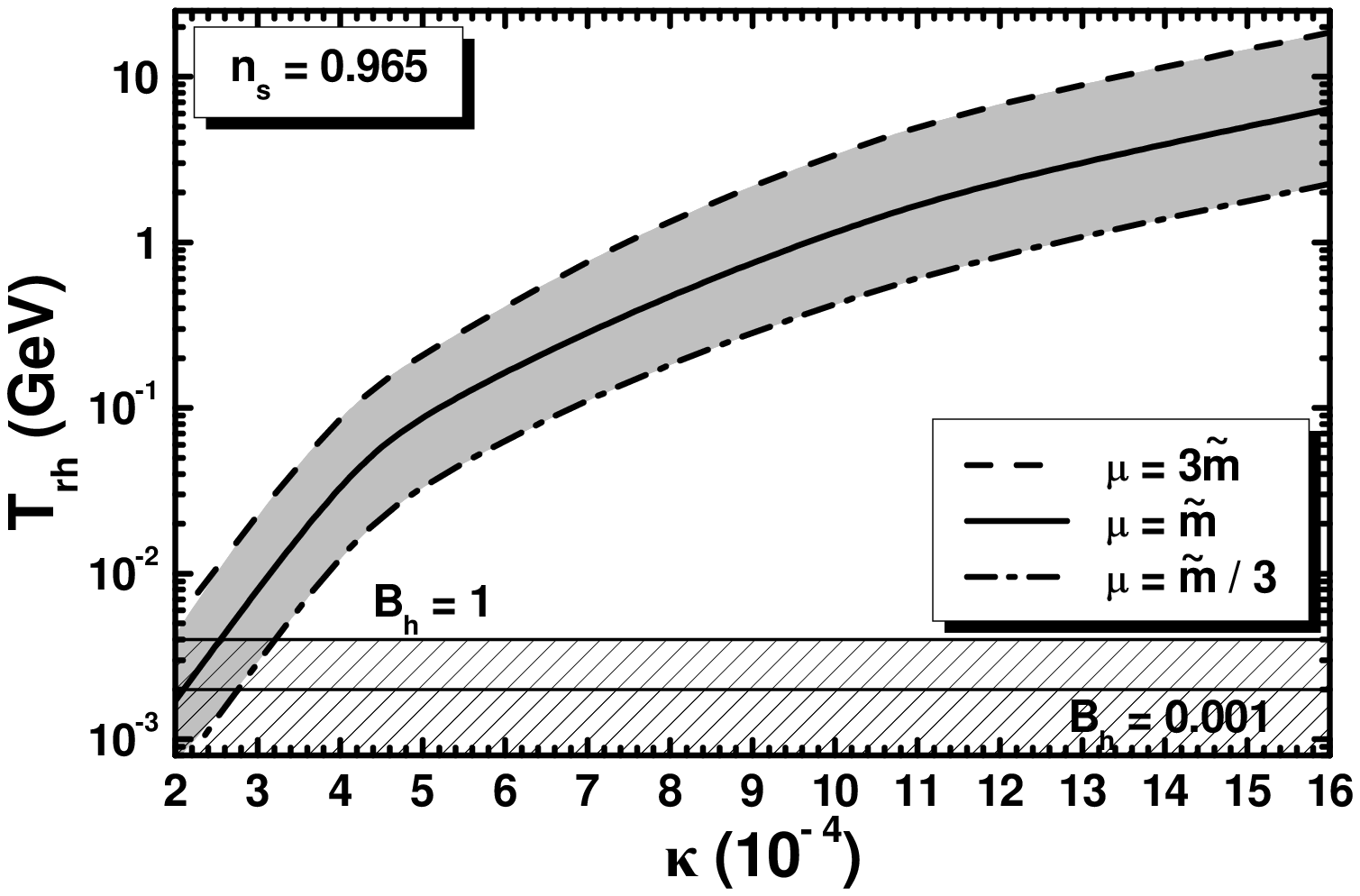}\eec
\caption{\sl Allowed strip in the $\kp-\Trh$ plane compatible with
the inflationary requirements in \Sref{fhi3} for $\ns=0.965$. We
take $\nu=7/8$ and $\mu=\mss$ (solid line), $\mu=\mss/3$
(dot-dashed line) or $\mu=3\mss$ (dashed line). The BBN lower
bounds on $\Trh$ for hadronic branching ratios $\br=1$ and $0.001$
are also depicted by two thin lines.}\label{figrh}
\end{figure}

If we enforce the compatibility between theoretical and
observational values of light element abundances predicted by BBN
we achieve \cref{nsref} a lower bound on $\Trh$ which, for large
$\mz\sim0.1~\PeV$, entails
\beq \Trh\geq4.1~\MeV~~\mbox{for}~~\br=1~~\mbox{and}~~
\Trh\geq2.1~\MeV~~\mbox{for}~~\br=10^{-3},\label{tns}\eeq
where $\br$ is the hadronic branching ratio. The bound above is a
little softened for larger $\mz$ values.

Taking $\kp$ and $\mz$ values allowed by the inflationary part of
our model in \Sref{fhi4}, we evaluate $\Trh$ as a function of
$\kp$ and delineate the regions allowed by the BBN constraints in
\Eref{tns} -- see \Sref{fhi3} below. The results of a such
computation are displayed in \Fref{figrh}, where we design allowed
contours in $\kp-\Trh$ plane for $\nu=7/8$. This is an
intermediate value in the selected here margin $(3/4-1)$. The
boundary curves of the allowed region correspond to $\mu=\mss/3$
or $\lm=0.22$ (dot-dashed line) and $\mu=3\mss$ or $\lm=1.96$
(dashed line) whereas the solid line is obtained for $\mu=\mss$ or
$\lm=0.65$. Note that the relation between $\lm$ and $\mu/\mss$ is
given in \Eref{mssi}.  We see that there is an ample parameter
space consistent with the BBN bounds in \Eref{tns} depicted with
two horizontal lines. Since the inflationary requirements increase
the scale $m$ with $\kp$ and $m$ heavily influences $\mz$ and also
$\Trh$ -- see \Eref{Trh} -- $\Trh$ increases with $\kp$. The
maximal value of $\Trh$ for the selected $\nu$ is obtained for
$\mu=3\mss$ and is estimated to be
\beq T_{\rm rh}^{\max}\simeq19~\GeV \eeq
Obviously, decreasing $\mu$ below $\mss/3$, causes $\lm$, $\Gsn$
and $\Trh$ to decrease too, and the slice cut from the BBN bound
increases.

It is worth emphasizing that the reheating stage is not
instantaneous. In particular, the maximal temperature $\Tmax$
during this period is much larger than $\Trh$, which can be better
considered as the largest temperature of the radiation domination
\cite{kolb}. To be quantitative, $\Tmax$ can be calculated as
\cite{rh}
\beq \label{btmax}
\Tmax=\lf3/8\rg^{3/20}12^{1/4}3^{1/8}\mP^{1/4}\Gsn^{1/4}\rho_{z\rm
I}^{1/8}/g_{\rm rh*}^{1/4}\pi^{1/2}.\eeq
It is achieved \cite{rh} for $\btau=\btmax=0.39\ll\btrh$. The
large hierarchy between $\Tmax$ and $\Trh$ can be appreciated by
their numerical values displayed in \Tref{tab} for BPA and BPB. We
see that $\Tmax\sim1~\PeV$ whereas $\Trh\sim1~\GeV$. As a
consequence, the electroweak sphalerons are still operative and so
baryogenesis via leptogenesis is in principle feasible.

\begin{figure}[!t]\vspace*{-.25in}
\bec\includegraphics[width=60mm,angle=-90]{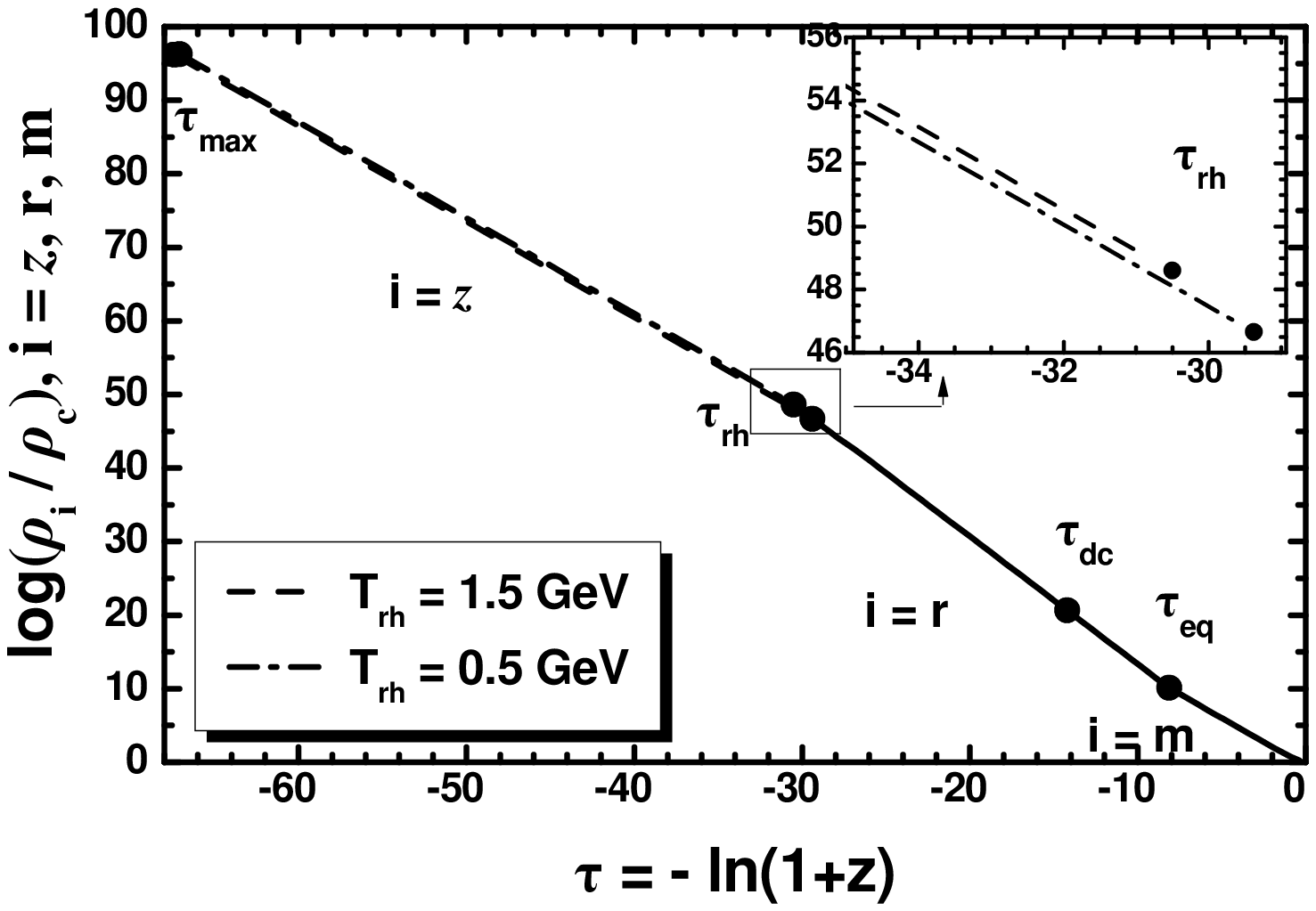}\eec
\caption{\sl Evolution of $\log\rho_i$ with $i=z$ -- dashed line
for $\Trh=1.5~\GeV$ and dot-dashed line for $\Trh=0.5~\GeV$ -- and
$i=$ r, m (solid lines) as a function of $\vtau=-\ln(1+\tz)$.
Shown are also \vtmax, \vtrh, \vteq and \vtmn for $\kp=0.001$ and
$\aS=25.3~\TeV$. }\label{figrhos}
\end{figure}

To obtain a more complete picture of our post-inflationary
cosmological setting we introduce the logarithmic time  $\vtau$
which is defined as a function of the redshift $\tz$ (not to be
confused with saxion field $z$) or the scale factor $R$ as
\cite{qcd}
\beq \vtau=-\ln (1+\tz)=\ln\lf R/R_0\rg,\label{dtau} \eeq
where the subscript $0$ is refereed hereafter to present-day
values. $\Tmax$ is achieved at $\vtmax$ which can be found from
\beq \vtmax=\btmax-\btrh+\vtrh~~\mbox{since}~~R_{\rm max}/R_0=\lf
R_{\rm max}/R_{I}\rg\cdot\lf R_{I}/R_{\rm rh}\rg\cdot\lf R_{\rm
rh}/R_{0}\rg \eeq
taking into account \eqs{btrh}{btmax}. For $\vtau\geq\vtrh$,
$\vtau$ can be found using the entropy conservation through the
relation
\beq \vtau= -\ln \lf g_{s*}/g_{0s*}\rg^{1/3}\lf T/T_0\rg \eeq
where $g_{s*}$ is the entropy effective number of degrees of
freedom at temperature $T$ and $T_0=2.35\cdot10^{-13}~\GeV$. Its
precise numerical values are evaluated by using the tables
included in public packages \cite{micro} and assuming the particle
spectrum of the MSSM. E.g, at BBN ($T=2~\MeV$) $\vtau_{\rm
BBN}\simeq-23$ and at the time of radiation-matter equidensity
$\vteq\simeq-8.1$ corresponding to $\tz_{\rm eq}=3387$. The values
of $\vtrh$ and $\vtmax$ are accumulated in \Tref{tab} for some
sample $\Trh$ and ratios $\mu/\mss$. We remark that $\vtmax$
values are similar for both $\Trh$ but $\vtrh$ is smaller for
larger $\Trh$.

The energy density of the oscillating $z$ condensate $\rho_z$,
radiation $\rho_{\rm r}$ and matter $\rho_{\rm m}$ may be
evaluated as follows\cite{qcd,rh}
\beq\label{rhos} \rho_z=\rho_{\rm
rh}e^{-3(\vtau-\vtrh)},~\rho_{\rm
r}=\frac{\pi^2}{30}g_{*}T^4~~\mbox{and}~~\rho_{\rm m}=\Omega_{{\rm
m0}}\rho_c e^{-3\vtau},
\end{equation}
where $\Omega_{\rm m0}=0.311$ and the energy density at reheating
is given by $\rho_{\rm rh}=\rho_{\rm r}(T=\Trh)$. Taking advantage
from the formulae above we illustrate in \Fref{figrhos} our
cosmological scenario. Namely, we present the evolution of the
various $\log\rho_i$ -- with $\rho_i$ normalized to $\rho_{\rm c}$
given in \Eref{rhoc} -- as functions of $\vtau$ keeping, for
simplicity, only the dominant component of the universal energy
density for each $\vtau$. As regards $\log\rho_i$ with $i=z$ we
employ the parameters of BPB in \Tref{tab} besides $\mu$ which is
$\mu=5\mss/4$ for $\Trh=1.5~\GeV$ (dashed line) or $\mu=2\mss/5$
for $\Trh=0.5~\GeV$ (dot-dashed line). We also draw with solid
lines $\log\rho_i$ with $i=$ r, m. From the plot it is apparent
that our scenario is distinguishable from the standard scenario
according to which the radiation domination commences after a high
$\Trh\sim1~\EeV$ corresponding to $\vtrh\sim-50$. We observe that
$\log\rho_z$ and $\log\rho_{\rm m}$ have the same slop since both
are proportional to $-3\vtau$ whereas $\log\rho_{\rm r}$ is more
stiff since it is proportional to $-4\vtau$. In the plot we also
localize the position of \vtmax, which is the same for both
$\Trh$'s, the two \vtrh\ values and the equidensity point \vteq.
Shown is also the $\vtau$ value which corresponds to the CS decay
\vtmn\ for $\srms=8$ -- see \Sref{css2}. It is located within the
radiation dominated era.

\section{Metastable CSs And An Early Matter Domination}\label{css}

The $U(1)_{B-L}$ breaking which occurs for $\sg\simeq\sgc$
produces a network of CSs which may be stable meta- or quasi-
stable. This network has the potential to undergo decay via the
Schwinger production of monopole-antimonopole pairs leading
thereby to the generation of a stochastic GW background. We below
compute the tension of these CSs in \Sref{css1} and their GW
spectrum in \Sref{css2} under the assumption that these are
metastable.


\subsection{CS Tension} \label{css1}

The dimensionless tension $\mcs$ of the $B-L$ CSs produced at the
end of FHI can be estimated by \cite{mfhi,buchbl}
\begin{equation} \label{mucs} \mcs \simeq
\frac12\lf\frac{M}{\mP}\rg^2\ecs(\rcs)~~\mbox{with}~~\ecs(\rcs)=\frac{2.4}{\ln(2/\rcs)}~~
\mbox{and}~~\rcs=\kappa^2/8g^2\leq10^{-2},\end{equation} where we
take into account that $(B-L)(\phc)=2$ -- cf. \cref{asfhi}. Here
$G=1/8\pi\mP^2$ is the Newton gravitational constant and
$g\simeq0.7$ is the gauge coupling constant at a scale close to
$M$. For the parameters in \Eref{res1} we find
\beq 0.59\lesssim\mcs/10^{-8}\lesssim9.2.\label{rescs}\eeq
To qualify the result above, we single out the cases:

\subparagraph{\ftn\sf (a)} If the CSs are stable, the range in
\Eref{rescs} is acceptable by the level of the CS contribution to
the observed anisotropies of CMB which is confined by \plk\
\cite{plcs0} in the range
\beq \mcs\lesssim 2.4\cdot 10^{-7}~~\mbox{at 95$\%$ c.l.}
\label{plcs} \eeq
On the other hand, the results of \Eref{rescs} are completely
excluded by the recent PTA bound which requires \cite{nano1}
\beq \mcs\lesssim 2\cdot 10^{-10}~~ \mbox{at 95$\%$ c.l.}
\label{ppta} \eeq

\subparagraph{\ftn\sf (b)} If the CSs are metastable, the
explanation \cite{meta1, nano1} of the recent \nano\ data
\cite{nano, pta} on stochastic GWs is possible for
\beq
M\gtrsim0.9~\YeV~~\mbox{and}~~\kp\gtrsim0.0003.\label{rescs1}\eeq
This is, because the obtained $\mcs$ values, through \Eref{mucs},
from the ranges above is confined in the range dictated by the
interpretation of the recent data \cite{nano1}
\beq  10^{-8}\lesssim  \mcs\lesssim 2.4\cdot
10^{-4}~~\mbox{for}~~8.2\gtrsim\sqrt{\rms}\gtrsim7.5~~ \mbox{at
95$\%$ c.l.}\label{kai} \eeq
where the metastability factor $\rms$ is the ratio of the monopole
mass squared to $\mu_{\rm cs}$. Since we do not investigate the
monopole formation in our work, the last restriction does not
impact on our parameters. Moreover, the GWs obtained from CSs have
to be consistent with the upper bound on their abundance $\ogw$
originating from the advanced \ligo\ third observing run
\cite{ligo}
\beq \ogw\lesssim 6.96\cdot10^{-9}~~\mbox{for}~~f=32~{\rm
Hz},\label{lvkb}\eeq
where $f$ is the frequency of the observation. At last, although
not relevant for our computation, let us mention for completeness
that $\ogw$ should be smaller than the limit on dark radiation
which is encoded in an upper limit on $\dN$ from BBN and CMB
observations \cite{bbngws}
\beq \ogw\lesssim
\frac{7}{8}\lf\frac{4}{11}\rg^{4/3}\Omega_{\gamma0}
h^2=5.6\cdot10^{-6}\dN~~\mbox{with}~~\dN\lesssim0.28~~\mbox{\rm at
$95\%$ c.l.},\label{gwnnb}\eeq
taking into account TT,TE,EE+lowE+lensing+BAO data sets
\cite{plcp}. Here, $\Omega_{\gamma0} h^2\simeq2.5\cdot10^{-5}$ is
the photon relic abundance at present and we assume that $\ogw$
develops a flat shape in a broad range of $f$ values so as the $f$
dependence in \Eref{gwnnb} to be very weak.


\subsection{GWs from Metastable CSs with Low Reheating}\label{css2}

We focus here on case {\ftn\sf (b)} of \Sref{css1} and compute the
spectrum of the GW produced by the CSs. The presence of the
long-lasting matter domination obtained in our set up due to the
$z$ oscillations after the end of FHI -- see \Sref{fhi} -- has
important ramifications in the shape of the spectrum of GW. This
is, because a decaying-particle-dominated era can be approximated
by a matter domination which leads to a spectral suppression at
relatively large frequencies \cite{wells,pillado}.

To verify this fact in the case of our model we compute the
emitted GW background at frequency $f$ following the standard
formula of \cref{pillado}
\begin{equation}\label{3}
\Omega_\text{GW}h^2=\frac{8\pi h^2}{3H_0^2}f (\mcs)^2
\sum_{\tk=1}^{\infty}
C_\tk(f)P_\tk~~\mbox{with}~~H_0=\sqrt{\rho_{\rm c}/{3\mP^2}}.
\end{equation}
%
%
Here $\rho_{\rm c}$ is given in \Eref{rhoc} and $P_\tk$ is the
power spectrum of GWs emitted by the $\tk^\text{th}$ harmonic of a
CS loop. Assuming cusps as the main source of GWs, $P_\tk$ is
given by \cite{pillado}
\begin{equation}
P_\tk\simeq{\Gm}/{\zeta({4}/{3})}\tk^{-4/3}~~\mbox{with}~~\Gm=50~~\mbox{and}~~\zeta({4}/{3})=3.6.
\end{equation}
In \Eref{3} $\ogw$ is expressed as a sum of the contributions from
an infinite number of normal modes. On technical ground, however,
we take a sum up to $\tk_{\rm max}\sim10^5$ to achieve a
sufficiently accurate result.

\begin{figure}[!t]\vspace*{-.16in}
\hspace*{-.19in}
\begin{minipage}{8in}
\epsfig{file=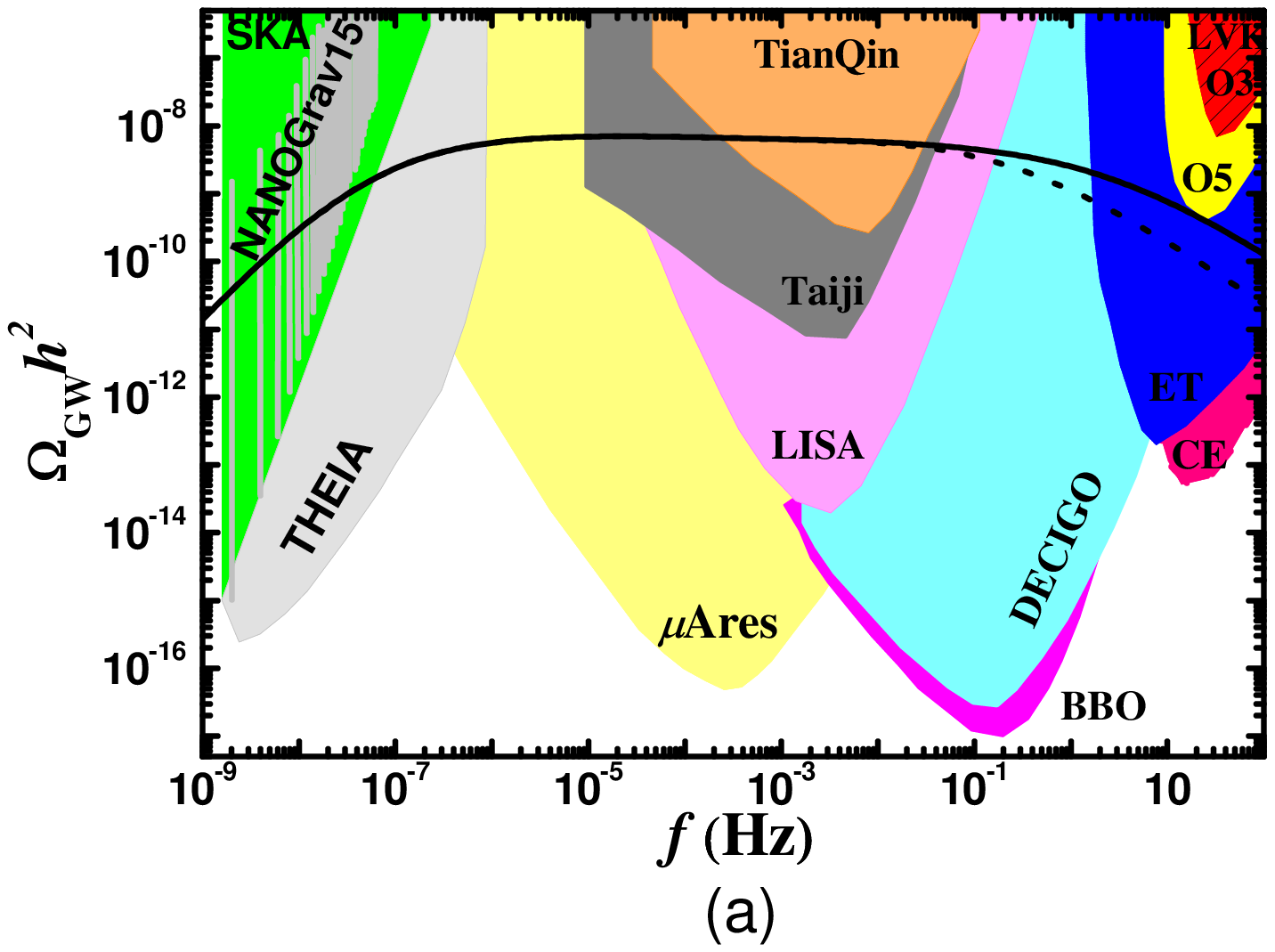,height=3.6in,angle=-90} \hspace*{-1.2cm}
\epsfig{file=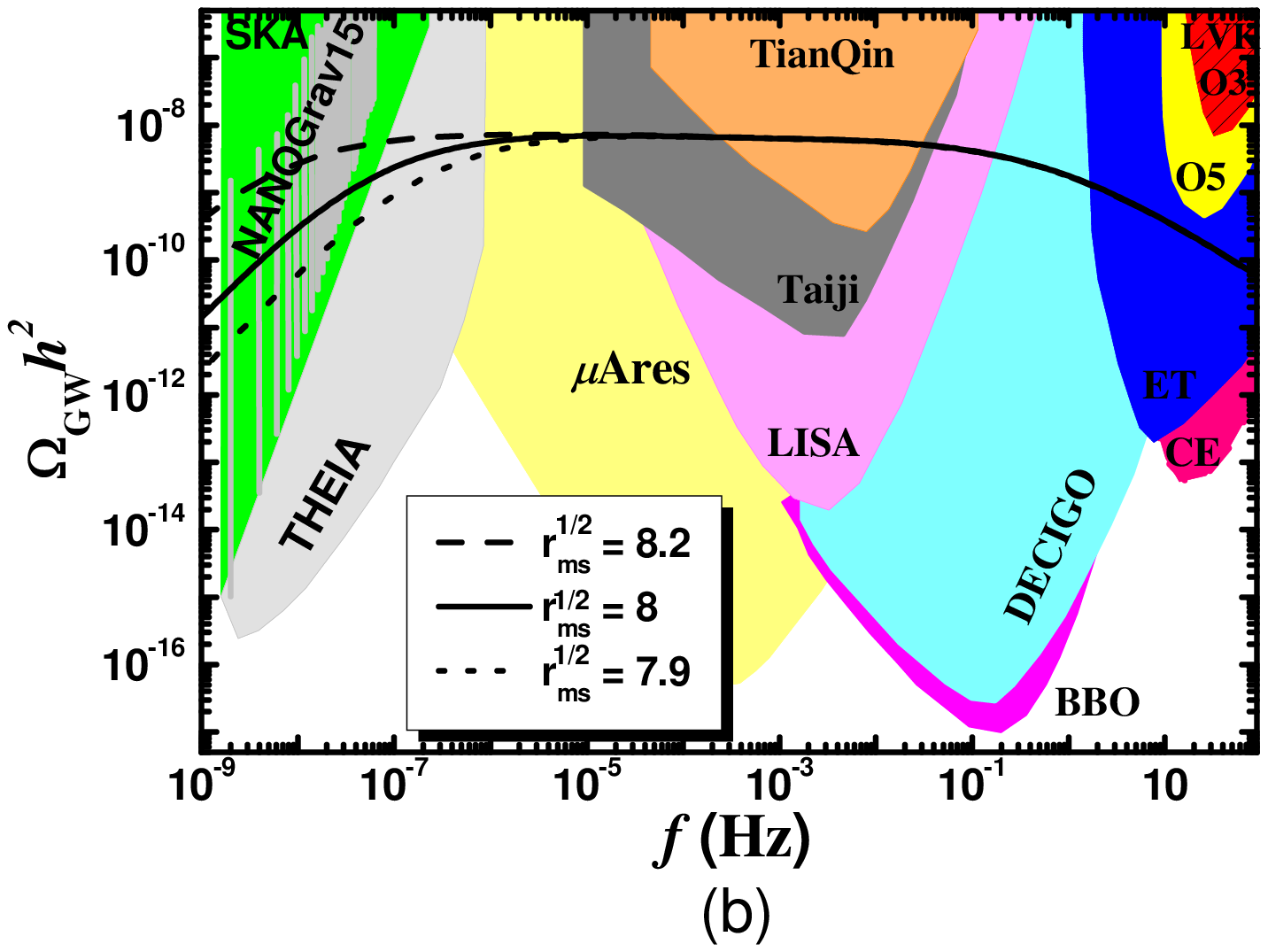,height=3.6in,angle=-90} \hfill
\end{minipage}
\caption{\sl GW spectra from $B-L$ metastable CSs for the inputs
of BPB and {\sffamily\ftn (a)} $\srms=8$  and $\mu=\mss/3$ (dotted
line) or $\mu=3\mss$ (solid line) {\sffamily\ftn (b)} $\mu=\mss$
and various metastability factors $\srms$ indicated in the plot.
The shaded areas in the background indicate the sensitivities of
the current -- i.e. NANOGrav \cite{nano} and LVK \cite{ligo} --
and future -- SKA \cite{ska}, THEIA \cite{thia}, $\mu$Ares
\cite{mares}, LISA \cite{lisa}, Taiji \cite{tj}, TianQin
\cite{tq}, BBO \cite{bbo}, DECIGO \cite{decig}, ET \cite{et} and
CE \cite{ce} -- experiments.}\label{figw}
\end{figure}


Following our cosmological scenario, the number of loops emitting
GWs, observed at a given frequency $f$, can be found from
\bea \nonumber C_\tk(f)&=&\frac{2\tk}{f^2}\lf
\int_{\vtmax}^{\vtrh}d\vtau e^{5\vtau} H_z n_{\rm m}\right.
+\int_{\vtrh}^{{\rm min}\{\vteq, \vtmn\}}d\vtau e^{5\vtau} H_{\rm
st} n_{\rm r}\nonumber \\ &+&\int_{\vteq}^{\vtmn}d\vtau e^{5\vtau}
H_{\rm st} n_{\rm rm}\Theta(\vtmn-\vteq)
+\left.\int_{\vteq}^{\vtmn}d\vtau e^{5\vtau} H_{\rm st} n_{\rm m}
\Theta(\vtmn-\vteq)\rg,\label{Ck}\eea
where we use as integration variable $\vtau$ -- cf. \cref{pillado}
-- taking into account \Eref{dtau} and $d\tz=-e^{-\vtau}d\vtau$.
Also \vtmax, \vtrh\ and \vteq\ are given in \Sref{rhs} and $\vtmn$
corresponds to the decay $\vtau$ of CS network, which is estimated
from
\beq
\vtmn=-\ln\Bigg(\left({70}/{H_0}\right)^{1/2}\left(\Gamma\Gamma_{\rm
dc}\mcs\rg^{1/4}+1\Bigg)~~\mbox{with}~~\Gamma_{\rm
dc}=4\mcs\mP^2e^{-\pi\rms}\eeq
the rate per unit length of the metastable CSs \cite{buch1}. For
the loop number density per unit length $n(\ell, t)$ -- with mass
dimension $4$ -- we adopt the expressions \cite{pillado}
\beqs\bea n_{\rm
r}(\ell,t)&=&\frac{0.18}{t^{3/2}\elg^{5/2}}\Theta(0.1
t-\ell),\label{nr}\\ n_{\rm
m}(\ell,t)&=&\frac{0.27-0.45(\ell/t)^{0.31}}{t^2\ell_{\Gamma}^{2}}\Theta(0.18
t-\ell),\\ n_{\rm rm}(\ell,t)&=&\frac{0.18t_{\rm
eq}^{1/2}}{t^2\elg^{5/2}}\Theta(0.09 t_\text{eq} - \elg-\ell),
\label{nm}\eea\eeqs
where the subscripts {\rm r} and {\rm m} are refereed to CSs
produced and radiating in a radiation- and a matter-dominated era
respectively whereas {\rm rm} means that the loops produced during
the radiation domination, but radiating during the matter
domination. Also $\elg$ is the initial length of a loop which has
length $\ell$ at a cosmic time $t$ and is given by
\beq \elg=\ell+\Gm\mcs t~~\mbox{with}~~ \Gm \simeq
50~~\mbox{and}~~\ell ={2\tk}e^\vtau/f. \eeq
Here $\Gm$ is related to the energy emission from CS \cite{wells}
and introduces some uncertainty in the computation. The Hubble
parameter during the standard cosmological evolution, $H_{\rm
st}$, and during the $z$ oscillation, $H_z$, can be found from the
formulas \cite{qcd, kolb,rh}
\begin{equation}\label{rhoqi}
H_{\rm st}={1\over\sqrt{3}\mP} \left(\Omega_\Lambda\rho_{\rm c0}
+\rho_{\rm m} +\rho_{{\rm
r}}\right)^{1/2}~~\mbox{and}~~H_z=\rho^{1/2}_z/\sqrt{3}\mP,\eeq
where the various $\rho_i$ are given in \Eref{rhos}. Lastly, the
cosmic time as a function of $\vtau$ is written as
\bea\nonumber
t(\vtau)&=&\int_{\vtmax}^{\vtau}\frac{d\vtau^\prime}{H_z}\Theta(\vtau-\vtmax)\Theta(\vtrh-\vtau)+
\int_{\vtmax}^{\vtau}\frac{d\vtau^\prime}{H_{\rm
st}}\Theta(\vtau-\vtrh)\\&\simeq&\begin{cases}
%
2/3H_z&\mbox{for}\>\>\>\vtmax \leq\vtau<\vtrh,\\
2/3(1+w)H_{\rm st}&\mbox{for}\>\>\>\vtau>\vtrh.\end{cases}
\label{tvtau}\eea
In particular, we have $w=1/3$ or $w=0$ for $\vtrh<\vtau\leq\vteq$
or $\vtau>\vteq$ respectively.

Armed with the formulae above we compute $\ogw$ for the GWs
produced from the CS formatted in our setting under the assumption
that these are metastable. Employing the inputs of BPB in
\Tref{tab} -- which yield $\mcs=6\cdot10^{-8}$ -- we obtain the
outputs displayed in \Fref{figw}. Namely, in \sFref{figw}{a} we
show $\ogw$ as a function of $f$ for $\srms=8$ and $\mu=\mss/3$
which yields $\Trh=0.4$ (dotted line) or $\mu=3\mss$ which results
to $\Trh=3.5$ (solid line). On the other hand, for the GW spectra
depicted in \sFref{figw}{b} we employ $\mu=\mss$ resulting to
$\Trh=1.2~\GeV$ and fix $\srms$ to $7.9$ (dotted line) $8$ (solid
line) and $8.2$ (dashed line) -- see \Eref{kai}. In both panels of
\Fref{figw} we see that the derived GW spectra can explain \nano\
data shown with gray almost vertical lines. We see though that, as
$\srms$ increases, the increase of $\ogw$ becomes sharper and
provide better fit to the observations. Also, in both panels the
shape of GW signal suffers a diminishment above a turning
frequency $f_{\rm rh}\sim0.03~\hz$ which enables us to satisfy
\Eref{lvkb} more comfortably than in the case with high reheating
-- cf. \cref{nasri,infl}. As $\Trh$ decreases, the reduction of
$\ogw$ becomes more drastic in accordance with the findings of
\cref{wells,pillado}. The plots also show examples of
sensitivities of possible future observatories \cite{ska, thia,
mares, lisa, tj, tq, bbo, decig, et, ce} which can test the
signals at various $f$ values. Needless to say, the bound in
\Eref{gwnnb} is not depicted since it is lies above the $\ogw$
values displayed in the plots.

It would be preferable to obtain larger $M$, and so $\mcs$, values
(e.g., $M\simeq10~\YeV$ yields $\mcs\simeq10^{-6}$) such that the
resulting $\ogw$ enters the dense part of the \nano\ favorable
region. This can be achieved -- see e.g. \cref{nasri,ahmed24} --
if we use $\theta=\pi$ and a next-to-minimal version \cite{hinova}
for $K_{\rm I}$. However, the generation of $\aS$ from $Z$ with
$\theta=0$ in \Eref{aSn} fixes the sign of the second term in
\Eref{cssb} and does not allow for the alternative arrangement
mentioned above. On the other hand, next-to-next-to-minimal
$K_{\rm I}$ may accommodate \cite{hinova, rlarge, ahmed24} such an
augmentation of the $M$ value without disturbing the $z$
stabilization during FHI -- see \Sref{fhi1}. In that case we
expect that $\aS$ would not be constrained by the requirements of
successful FHI and it could be selected so that $\Trh$ lies at the
convenient level which allows for the comfortable evasion of the
exclusion limits from the \ligo\ \cite{ligo} experiment. This
setting opens up an interesting interconnection between the GW
experiments and $\mss$.







\section{Predictions for the SUSY-Mass Scale}\label{susy}

The $\aS$ values displayed in Fig.~\ref{fig2} and \Eref{res1} --
which result from the observational constraints to FHI  -- give us
the opportunity to gain information about the mass scale of SUSY
particles through the determination of $\mss=\mgr$ -- see
\Eref{mssi}. This aim can be achieved by solving \Eref{aSn} w.r.t.
$m$ as follows
\beq
m\simeq\lf\frac{\aS}{2^{1+\nu}(2-\nu)}\rg^{(2-\nu)/2}\lf\frac{3\Hhi^2}{(1-\nu)\nu^2}\rg^{\nu/4},\label{maS}\eeq
where we take into account \Eref{veviz} and the fact that
$\vevi{z}/\mP\sim10^{-3}$. As a consequence, the analytic result
above approximates well the numerical one which is obtained by
extracting consistently $m$ as a function of $\kp$ and $M$
determined by the conditions in \eqs{Nhi}{Prob}. Note that an
iterative process has to be realized introducing a trial $m$ value
which allows us to use as input the form of $\Vhi$ in \Eref{vol}.
The aforementioned smallness of $\vevi{z}$ causes a diminishment
of $m$ compared to $\aS$ rendering therefore $\mss$ via
\eqs{mssi}{mgr} of the order of $\PeV$ scale. Indeed, for fixed
$\nu$, \Eref{maS} yields $m$ and then $\mgr$ , $\mz$ and $\mth$
can be easily estimated by \eqs{mgr}{mzth} whereas $\mss$ and
$\Trh$ by \eqs{mssi}{Trh}. Their numerical values for the BPs A
and B are presented in \Tref{tab}. The magnitude of the derived
$\mss$ values and the necessity for $\mu\sim\mss$, established in
\Sref{rhs} hints towards the high-scale MSSM.

\begin{figure}[!t]\vspace*{-.25in}
\bec\includegraphics[width=60mm,angle=-90]{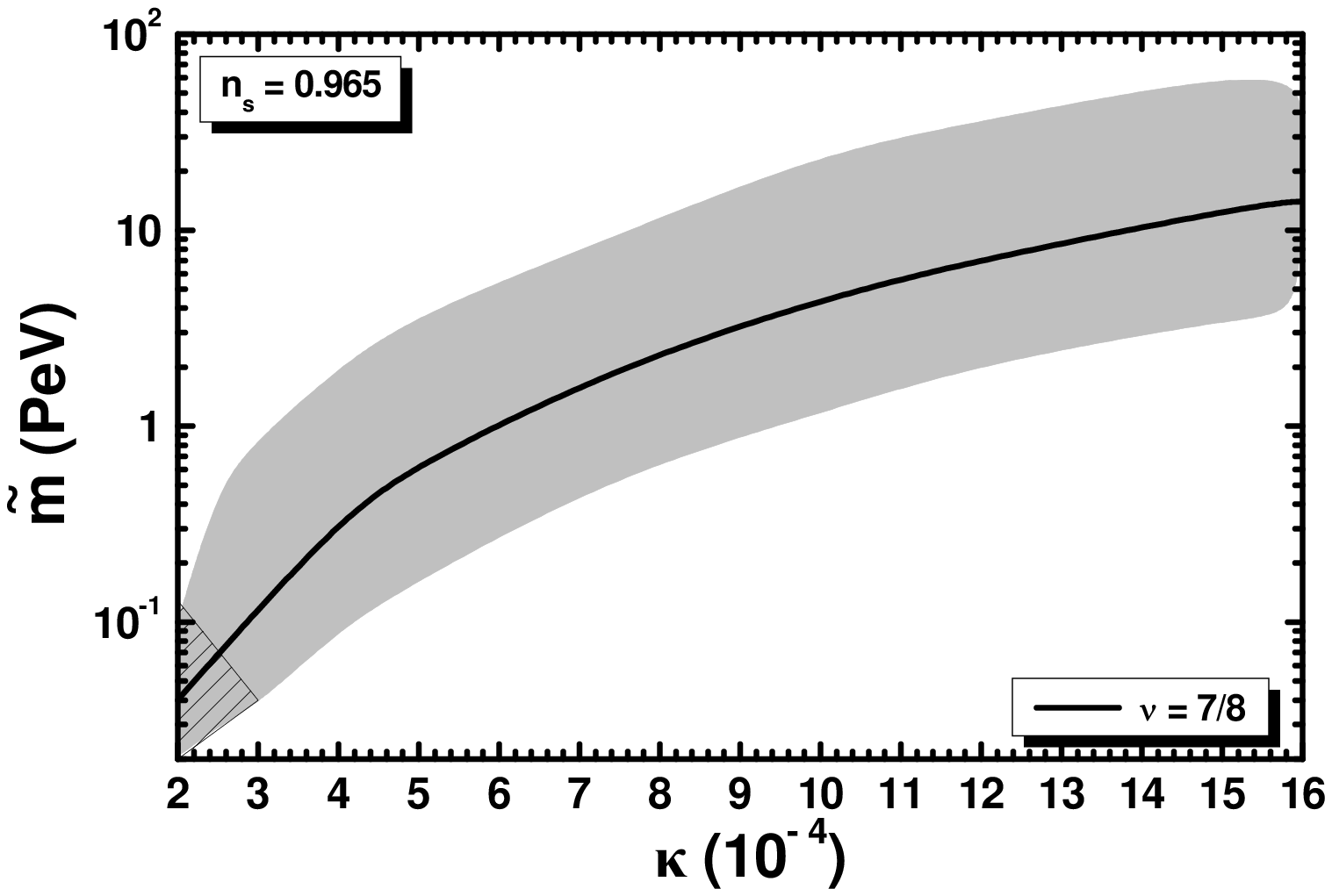}\eec
\caption{\sl Allowed by Eqs.~(\ref{Nhi}) and (\ref{Prob}) region
in the $\kp-\mss$ plane for $\mss/3\leq \mu\leq3\mss$, $\ns=0.965$
and $3/4<\nu<1$. The allowed contour for $\nu=7/8$ are also
depicted. Hatched is the region excluded by BBN for
$\br=0.001$.}\label{fig4}
\end{figure}

To highlight numerically our expectations above, we display the
allowed (gray shaded) region in the $\kp-\mss$ plane fixing
$\ns=0.965$ and varying $\nu$ and $\mu$ within their possible
respective margins $(0.75-1)$ and $(1/3-3)\mss$ -- see
\Fref{fig4}. Along the solid line we set $\nu=7/8$. From
\Eref{maS} we can convince ourselves that the lower boundary curve
of the displayed region is obtained for $\nu\simeq0.751$, whereas
the upper one corresponds to $\nu\simeq0.99$. Assuming also
$\mu=\mss$ we can determine the slice of the area which can be
excluded due to the BBN bound in \Eref{tns}. In all, we find that
$\mss$ turns out to be confined in the range
\beq 1.2\lesssim \aS/\TeV\lesssim
100~~\mbox{and}~~0.34\lesssim\mss/\PeV\lesssim
13.6,\label{mss1}\eeq
whereas $T_{\rm rh}^{\max}\simeq71~\GeV$ attained for $\mu=3\mss$
and $\nu\simeq0.99$. The derived allowed margin of $\mss$ and the
employed $\mu$ values render our proposal compatible with the mass
of the Higgs boson discovered in LHC, if we adopt as a low energy
effective theory the high-scale version of MSSM \cite{strumia}.
Indeed, within high-scale SUSY, updated analysis requires
\cite{strumia}
\beq 3~\TeV\lesssim\mss\lesssim0.3~\ZeV,\label{highb} \eeq
for degenerate sparticle spectrum, $\mss/3\leq\mu\leq3\mss$,
$1\leq\tan\beta\leq50$ and varying the stop mixing. On the other
hand, our setting does not fit well with natural \cite{baerh} or
split \cite{strumia} SUSY which assume $\mu\ll\mss$.

\newpage

\section{Conclusions}\label{con}

We analyzed the implementation of FHI together with the SUSY
breaking and the CS formation in a B-L extension of MSSM. We
adopted the super- and \Kaa\ potentials in \eqs{Who}{Kho} applying
an approximate $R$ symmetry. The model offers the following
interesting achievements:

\begin{itemize}

\item Observationally acceptable FHI of hilltop-type adjusting the
tadpole parameter $\aS$ and the $B-L$ breaking scale $M$ in the
ranges of \Eref{res1}.

\item A prediction for the SUSY-mass scale $\mss$ which turns out
to be of the order of PeV.

\item An interpretation of the DE problem without extensive
tuning. We obtain $\lambda\sim10^{-12}$ in \Eref{wgh}.

\item Explantation of the $\mu$ term of MSSM with $|\mu|\sim\mgr$
-- see \Eref{dK} -- by invoking a appropriate modification of the
Giudice-Masiero mechanism.

\item  Reheating generated due to the domination of the sgoldstino
condensate after the end of FHI; since $\mu$ is of the order PeV,
the resulting $\Trh$ is higher than its lower bound from BBN.

\item $B-L$ CSs which, if metastable, explain rather well the
recent observations on GWs.  A characteristic of the obtained
spectra is their suppression at relatively large frequencies
($f>0.1~{\rm Hz}$) which is due to the long-lasting matter
domination caused by the sgolstino oscillations.

\item  Generation of neutrino masses via the type I see-saw
mechanism supported by $W_{\rm MD}$ and $K_{\rm D}$ in
\eqs{wmd}{kd}.

\end{itemize}


A potential shortcoming of our proposal is that baryogenesis is
made difficult due to the low reheat temperature. Non-thermal
leptogenesis is not operative since the sgoldstino (with mass of
order $1~\PeV$) is lighter than $N_i^c$ which acquire mass of
order $1~\ZeV$ -- cf. \cref{mfhi,univ} -- and so the direct decay
of $z$ into $N_i^c$ is forbidden. However, there are extensions of
MSSM \cite{allahbau} where the late decay of the sgoldstino may
generate non-thermally the baryon asymmetry of the universe.
Alternatively, this problem may be overcome applying improved
attempts \cite{bau} based on the idea of cold electroweak
baryogenesis \cite{cbau}. As regards the CDM, the candidacy of
lightest neutralino has to be investigated thoroughly solving
precisely the relevant Bolzmann equations, as in \cref{rh,
kolbdm}. In case that the abundance is low we can open up slightly
the decay channel of $z$ into $\Gr$ so that we obtain a
controllable production of neutralinos. Another aspect is the fate
of $R$ axion which remains stable throughout our setting
\cite{Raxiondm}.

As regards naturalness, it is a puzzle why the SUSY should appear
at such scale, which is higher than the electroweak scale making
the fine-tuning severe, while it is much smaller than the
fundamental energy scale such as the GUT or Planck scales. Our
scenario provides a possible solution to this issue: this may be
due to the inflationary selection. Namely, the apparent
fine-tuning could be a result of combination of the FHI and a bias
toward high-scale SUSY in the landscape.

\subparagraph{\bfseries\scshape Acknowledgments: } I would like to
thank R. Kolb and R. Maji for useful discussions. This research
work was supported by the Hellenic Foundation for Research and
Innovation (H.F.R.I.) under the ``First Call for H.F.R.I. Research
Projects to support Faculty members and Researchers and the
procurement of high-cost research equipment grant'' (Project
Number: 2251).

\subparagraph{\small\bfseries\scshape Dedication: } {\small It is
with deep sadness that I dedicate this work to the memory of my
Ph. D advisor, Prof. George Lazarides. He has been a giving mentor
and an invaluable collaborator who has constantly supported me
over the past thirty years.}



\newcommand\mtta[4]{\mbox{
$\llgm\bem #1 &#2 \cr #3& #4\eem\rrgm$}}

\appendix{Dirac and Majorana Masses to Neutrinos}\label{app}

\paragraph{\hspace*{.25cm}} We check here if the presence of $K_{\rm D}$ in \Eref{Kho}
-- which is absent in other similar settings \cite{mfhi,univ} --
has some impact on the derivation of the neutrino masses. This is
because $K_{\rm D}$ in \Eref{kd} contributes to Dirac neutrino
masses according to the formulae of \cref{tamv}. In our case the
non-vanishing contributions are
\begin{align} \nonumber m_{\al\bt}&=\frac12\lf
\vev{\partial^2_{\al\bt}W_{\rm MD}}-\vev{K_{\rm
H}^{ZZ^*}\partial^3_{\al\bt Z^*}K_{\rm HO}\partial_ZW_{\rm
H}}\rg\\ &-\frac{\mgr}{2}\lf \vev{K_{\rm
H}^{ZZ^*}\partial^3_{\al\bt Z^*}K_{\rm HO}\partial_Z K_{\rm
H}}-\vev{\partial^2_{\al\bt} K_{\rm HO}}\rg, \label{mab}
\end{align}
where $\partial_\al:=\partial/\partial{Y^\al}$,
$\partial_Z:=\partial/\partial{Z}$ and we define
\beq K_{\rm HO}=K_{\rm H}+K_\mu+K_{\rm
D}+|Y^\al|^2~~\mbox{with}~~Y^\al=\hu, \hd, \phc, \phcb,
L~~\mbox{and}~~N^c.\eeq
If we confine ourselves to the case of the third generation we
obtain the following mass matrix for the neutrino masses
\beq  \mtta{0}{\mD[K]+\mD[W]}{\mD[K]+\mD[W]}{\mM},\eeq
where the various contributions read
\bea &&\mD[W]= h_{N}\vev{\hu}/2~~\&~~ \mM=(2\nu)^\nu M/3^{\nu/2} \\
&&\mD[K]=\frac{4^nm}{3^n\mP}\ld_{\rm D} \nu^{2\nu}\vev{\hu}\om\lf
\om^{N/2-2}\lf\frac12-\nu+\frac23\nu^2\rg-\frac38\rg, \eea
from which those in the first line come from the terms of
\Eref{wmd} whereas this in the second line originates from
\Eref{kd}. Note that $\om$ is given in \Eref{vcc}. For the inputs
of BPB in \Tref{tab} we obtain from the well-known (type I)
see-saw formula -- see e.g. \cref{univ} -- for the mass of the
third generation neutrino, $\nu_3$, and $N^c$, $N^c_3$,
respectively
\beq m_{3\nu}\simeq\mD[W]^2/M\simeq0.05~\eV~~\mbox{and}~~
M_{3N^c}\simeq63~\ZeV~~
\mbox{if}~~h_{3N}=0.5~~\mbox{and}~~\ld_{\rm 3N^c}=0.03.\eeq
The obtained $m_{3\nu}$ is phenomenologically acceptable if we
assume that the three active $\nu_i$'s have the normal mass
hierarchy so that $m_{3\nu}$ is equal to the mass induced by
atmospheric $\nu_i$ experiments. For the sample values above we
remark that $\mD[W]\sim40~\GeV$ dominates over
$\mD[K]\sim0.01~\eV$. However, if the terms in $W_{\rm DM}$ were
prohibited then $\mD[K]$ could naturally account for the neutrino
masses thanks to the high $\mgr\sim\PeV$ \cite{wellspev}. Indeed,
the resulting $\mD[K]$ takes the correct value for $\ld_{\rm
D}\simeq2.05$.

In conclusion, our model assures acceptable neutrino masses from
the superpotential term in \Eref{wmd} without sizable contribution
from the terms of \Eref{kd} emerging in $K$.

\newpage

\appendix{Abbreviations}\label{app1}
\paragraph{\hspace*{.25cm}} For further clarity we list here the abbreviations used in this manuscript.
\\ \\
\begin{tabular}{@{}lcl}
BAO &:& Baryon Acoustic Oscillations\\
BBN &:& Big Bang Nucleosynthesis\\
BPA, B  &:& Benchmark Point A, B\\
CDM &:& Cold Dark Matter \\
CMB &:& Cosmic Microwave Background\\
CS &:&  Cosmic String\\
DE &:&  Dark Energy\\
dS &:& de Sitter\\
FHI &:&  F-term Hybrid Inflation\\
GUT &:&  Grand Unified Theory \\
GW &:&  Gravitational Wave\\
HS &:&  Hidden Sector\\
LVK &:&  LIGO-VIRGO-KAGRA\\
IS &:&  Inflationary Sector\\
MSSM &:&  Minimal SUSY SM \\
NANOGrav15 &:& NANOGrav 15-years
results \\
RCs &:&  Radiative Corrections\\
SM &:&  Standard Model \\
SUSY &:&  Supersymmetry\\
SUGRA &:&  Supergravity\\
v.e.v &:& vacuum expectation value \\
w.r.t &:& with respect to
\end{tabular}

\def\ijmp#1#2#3{{\sl Int. Jour. Mod. Phys.}
{\bf #1},~#3~(#2)}
\def\plb#1#2#3{{\sl Phys. Lett. B }{\bf #1}, #3 (#2)}
\def\prl#1#2#3{{\sl Phys. Rev. Lett.}
{\bf #1},~#3~(#2)}
\def\rmp#1#2#3{{Rev. Mod. Phys.}
{\bf #1},~#3~(#2)}
\def\prep#1#2#3{{\sl Phys. Rep. }{\bf #1}, #3 (#2)}
\def\prd#1#2#3{{\sl Phys. Rev. D }{\bf #1}, #3 (#2)}
\def\npb#1#2#3{{\sl Nucl. Phys. }{\bf B#1}, #3 (#2)}
\def\npps#1#2#3{{Nucl. Phys. B (Proc. Sup.)}
{\bf #1},~#3~(#2)}
\def\mpl#1#2#3{{Mod. Phys. Lett.}
{\bf #1},~#3~(#2)}
\def\jetp#1#2#3{{JETP Lett. }{\bf #1}, #3 (#2)}
\def\app#1#2#3{{Acta Phys. Polon.}
{\bf #1},~#3~(#2)}
\def\ptp#1#2#3{{Prog. Theor. Phys.}
{\bf #1},~#3~(#2)}
\def\n#1#2#3{{Nature }{\bf #1},~#3~(#2)}
\def\apj#1#2#3{{Astrophys. J.}
{\bf #1},~#3~(#2)}
\def\mnras#1#2#3{{MNRAS }{\bf #1},~#3~(#2)}
\def\grg#1#2#3{{Gen. Rel. Grav.}
{\bf #1},~#3~(#2)}
\def\s#1#2#3{{Science }{\bf #1},~#3~(#2)}
\def\ibid#1#2#3{{\it ibid. }{\bf #1},~#3~(#2)}
\def\cpc#1#2#3{{Comput. Phys. Commun.}
{\bf #1},~#3~(#2)}
\def\astp#1#2#3{{Astropart. Phys.}
{\bf #1},~#3~(#2)}
\def\epjc#1#2#3{{Eur. Phys. J. C}
{\bf #1},~#3~(#2)}
\def\jhep#1#2#3{{\sl J. High Energy Phys.}
{\bf #1}, #3 (#2)}
\newcommand\njp[3]{{\sl New.\ J.\ Phys.\ }{\bf #1}, #3 (#2)}
\def\prdn#1#2#3#4{{\sl Phys. Rev. D }{\bf #1}, no. #4, #3 (#2)}
\def\jcapn#1#2#3#4{{\sl J. Cosmol. Astropart.
Phys. }{\bf #1}, no. #4, #3 (#2)}
\def\epjcn#1#2#3#4{{\sl Eur. Phys. J. C }{\bf #1}, no. #4, #3 (#2)}


\begin{thebibliography}{99}

\bibitem{susyhybrid} G.R. Dvali, Q. Shafi and R.K. Schaefer, {\it Large
scale structure and supersymmetric inflation without fine tuning},
\prl{73}{1994}{1886} [{\ftn\tt hep-ph/9406319}].


\bibitem{hinova} C.~Pallis,
{\it Reducing the spectral index in F-term hybrid inflation,}
edited by T.P. Harrison and R.N. Gonzales (Nova Science Publishers
Inc., New York, 2008) [\arxiv{0710.3074}]; R. Armillis and C.
Pallis, {\it Implementing Hilltop F-term Hybrid Inflation in
Supergravity}, {\sl ``Recent Advances in Cosmology''}, edited by
A. Travena and B. Soren (Nova Science Publishers Inc., New York,
2013)  [\arxiv{1211.4011}].


\bibitem{lectures} G. Lazarides, {\it Inflationary cosmology}, {\sl Lect. Notes Phys. }{\bf 592}, 351 (2002)
[{\ftn\tt hep-ph/0111328}]; \\ G. Lazarides, {\it Basics of
inflationary cosmology}, {\sl J. Phys. Conf. Ser.} {\bf 53}, 528
(2006) [{\ftn\tt hep-ph/0607032}].





\bibitem{plin} Y.~Akrami {\it et al.} [\plk\ Collaboration], {\it Planck
2018 results. X. Constraints on inflation}, {\sl Astron.
Astrophys. }\textbf{641}, A10 (2020) [\arxiv{1807.06211}].



\bibitem{pana} C. Panagiotakopoulos, {\it Hybrid inflation in supergravity
with $(SU(1, 1)/ U(1))^m$ K\"ahler manifolds}, {\sl Phys. Lett. B}
{\bf 459}, 473 (1999) [{\ftn\tt hep-ph/9904284}];
C.~Panagiotakopoulos, {\it Realizations of hybrid inflation in
supergravity with natural initial conditions}, {\sl Phys.\ Rev.\
D} {\bf 71}, 063516 (2005) [\hepph{0411143}].

\bibitem{gpp} M. Bastero-Gil, S.F. King and Q. Shafi, {\it Supersymmetric Hybrid
Inflation with Non-Minimal K\"ahler potential},
\plb{651}{2007}{345} [{\ftn\tt hep-ph/0604198}]; B. Garbrecht, C.
Pallis and A. Pilatsis, {\it Anatomy of F(D)-Term Hybrid
Inflation}, \jhep{12}{2006}{038} [{\ftn\tt hep-ph/0605264}]; M.U.
Rehman, V.N.~\c{S}eno\u{g}uz and Q.~Shafi, {\it Supersymmetric And
Smooth Hybrid Inflation In The Light Of WMAP3}, {\sl Phys. Rev. D
}{\bf 75}, 043522 (2007) [{\tt\ftn hep-ph/0612023}].

\bibitem{kelar} C.~Pallis, {\it K\"ahler Potentials for Hilltop F-Term Hybrid Inflation},
\jcap{04}{2009}{024} [\arxiv{0902.0334}].

\bibitem{rlarge} M.U.~Rehman, Q.~Shafi and J.R.~Wickman, {\it Observable Gravity Waves
from Supersymmetric Hybrid Inflation II}, {\sl Phys.\ Rev.\ D}
{\bf 83}, 067304 (2011) [\arxiv{1012.0309}]; M.~Civiletti,
C.~Pallis and Q.~Shafi, {\it Upper Bound on the Tensor-to-Scalar
Ratio in GUT-Scale Supersymmetric Hybrid Inflation},
\plb{733}{2014}{276} [\arxiv{1402.6254}].


\bibitem{sstad} V.N. \c{S}eno\u{g}uz and Q. Shafi, {\it Reheat temperature
in supersymmetric hybrid inflation models}, {\sl Phys.\ Rev.\
D}~{\bf 71}, 043514 (2005) [{\ftn\tt hep-ph/0412102}];
M.U.~Rehman, Q.~Shafi and J.R.~Wickman, {\it Supersymmetric Hybrid
Inflation Redux} {\sl Phys.\ Lett.\ B} {\bf 683}, 191 (2010)
[\arxiv{0908.3896}]; M.U.~Rehman, Q.~Shafi and J.R.~Wickman, {\it
Minimal Supersymmetric Hybrid Inflation, Flipped SU(5) and Proton
Decay}, {\sl Phys.\ Lett.\ B }{\bf 688}, 75 (2010) [{\tt\ftn
arXiv:0912.4737}]; K.~Nakayama \etal, {\it Constraint on the
gravitino mass in hybrid inflation} \jcap{12}{2010}{010}
[\arxiv{1007.5152}].

\bibitem{mfhi} C. Pallis and Q. Shafi, {\it Update on Minimal
Supersymmetric Hybrid Inflation in Light of PLANCK}, {\sl Phys.
Lett. B }\textbf{725}, 327 (2013) [\arxiv{1304.5202}].

\bibitem{kaihi} W. Buchm\"uller, V. Domcke, K. Kamada and K. Schmitz, {\it Hybrid
Inflation in the Complex Plane,} \jcap{07}{2014}{054}
[\arxiv{1404.1832}].
%

\bibitem{ahmed24} W.~Ahmed and S.~Raza, {\it Supersymmetric Hybrid Inflation in
Light of CMB Experiments and Swampland Conjectures,}
\arxiv{2401.02168}.

\bibitem{split} C.~Pallis and Q.~Shafi, {\it From Hybrid to Quadratic Inflation With
High-Scale Supersymmetry Breaking}, {\sl Phys. Lett. B}
\textbf{736}, 261 (2014) [\arxiv{1405.7645}].


\bibitem{rachel} R.~Jeannerot, J.~Rocher and M.~Sakellariadou,
{\it How generic is cosmic string formation in SUSY GUTs,} {\sl
Phys. Rev. D }\textbf{68}, 103514 (2003) [\hepph{0308134}].


\bibitem{univ} C.~Pallis, \textit{Gravitational Waves, $\mu$ Term
\& Leptogenesis from $B-L$ Higgs Inflation in Supergravity}, {\sl
Universe }\textbf{4}, no.1, 13 (2018) [\arxiv{1710.05759}].


\bibitem{pta} J. Antoniadis \etal\ [EPTA Collaboration]  {\it The second data release from the
European Pulsar Timing Array - III. Search for gravitational wave
signals,} {\sl Astron. Astrophys. }{\bf 678}, A50 (2023)
[\arxiv{2306.16214}]; D.J. Reardon \etal, {\it Search for an
Isotropic Gravitational-wave Background with the Parkes Pulsar
Timing Array}, {\sl Astrophys. J. Lett. }\textbf{951}, no.~1, L6
(2023) [\arxiv{2306.16215}]; H. Xu \etal, {\it Searching for the
Nano-Hertz Stochastic Gravitational Wave Background with the
Chinese Pulsar Timing Array Data Release I}, {\sl Res. Astron.
Astrophys. }{\bf 23}, no.~7, 075024 (2023) [\arxiv{2306.16216}].


\bibitem{nano} G. Agazie \etal\ [NANOGrav Collaboration], {\it The NANOGrav 15 yr
Data Set: Evidence for a Gravitational-wave Background}, {\sl
Astrophys. J. Lett. }{\bf 951}, no.~1, L8  (2023)
[\arxiv{2306.16213}].

\bibitem{nano1} A. Afzal \etal\ [NANOGrav Collaboration], {\it The
NANOGrav 15 yr Data Set: Search for Signals from New Physics},
{\sl Astrophys. J. Lett. }\textbf{951}, no.~1, L11 (2023)
\arxiv{2306.16219}.


\bibitem{buchfhi}  W. Buchm\"uller, {\it Metastable strings and dumbbells in
supersymmetric hybrid inflation}, \jhep{04}{2021}{168}
[\arxiv{2102.08923}].


\bibitem{so10} R. Maji, W.-I. Park and Q. Shafi, {\it Gravitational waves from walls bounded by strings in SO(10)
model of pseudo-Goldstone dark matter}, {\sl Phys. Lett. B} {\bf
845} 138127 (2023) [2305.11775]; S. Antusch, K. Hinze, S. Saad and
J. Steiner, {\it Singling out $SO(10)$ GUT models using recent PTA
results} \prdn{108}{2023}{095053}{9} [\arxiv{2307.04595}]; B. Fu,
S.F. King, L. Marsili, S. Pascoli, J. Turner, and Y.-L. Zhou, {\it
Testing Realistic $SO(10)$ SUSY GUTs with Proton Decay and
Gravitational Waves}, \arxiv{2308.05799}.

\bibitem{leont} S.F.~King, G.K.~Leontaris and Y.L.~Zhou, {\it Flipped SU(5):
unification, proton decay, fermion masses and gravitational
waves,} \jhep{03}{2024}{006}  [\arxiv{2311.11857}].

\bibitem{nasri} W.~Ahmed, M.~Junaid, S.~Nasri and U.~Zubair,
{\it Constraining the cosmic strings gravitational wave spectra in
no-scale inflation with viable gravitino dark matter and
nonthermal leptogenesis,} {\sl Phys. Rev. D }\textbf{105}, no.11,
115008 (2022) [\arxiv{2202.06216}]; W.~Ahmed, T.A.~Chowdhury,
S.~Nasri and S.~Saad, {\it Gravitational waves from metastable
cosmic strings in Pati-Salam model in light of new pulsar timing
array data,} {\sl Phys. Rev. D }{\bf 109} (2024) 015008
[\arxiv{2308.13248}].

\bibitem{su5} W. Ahmed, M.U. Rehman and U. Zubair, {\it Probing Stochastic
Gravitational Wave Background from $SU(5)\times U(1)$ Strings in
Light of NANOGrav 15-Year Data,} \jcap{01}{2024}{049}
[\arxiv{2308.09125}].

\bibitem{infl} G.~Lazarides, R.~Maji, A.~Moursy and Q.~Shafi, {\it Inflation,
superheavy metastable strings and gravitational waves in
non-supersymmetric flipped $SU(5)$} 
[\arxiv{2308.07094}]; A.~Afzal, M.~Mehmood, M.U.~Rehman and
Q.~Shafi, {\it Supersymmetric hybrid inflation and metastable
cosmic strings in $SU(4)_{\rm c}\times SU(2)_{\rm L}\times
U(1)_{\rm R}$}, \arxiv{2308.11410}.



\bibitem{meta1} W.~Buchm\"uller, V.~Domcke and K.~Schmitz,
{\it From NANOGrav to LIGO with metastable cosmic strings,} {\sl
Phys. Lett. B }\textbf{811}, 135914 (2020) [\arxiv{2009.10649}];
W. Buchm\"uller, V. Domcke and K. Schmitz, {\it Metastable cosmic
strings,} \jcap{11}{2023}{020} [\arxiv{2307.04691}].


\bibitem{quasi} G. Lazarides, R. Maji and Q. Shafi, {\it Gravitational waves
from quasi-stable strings}, \jcap{08}{2022}{042}
[\arxiv{2203.11204}]; G.~Lazarides, R.~Maji and Q.~Shafi, {\it
Superheavy quasistable strings and walls bounded by strings in the
light of NANOGrav 15~year data,} {\sl Phys. Rev. D }\textbf{108},
no.9, 095041 (2023) [\arxiv{2306.17788}].



\bibitem{buchbl} W. Buchm\"uller, V. Domcke and K. Schmitz, {\it Spontaneous
B-L Breaking as the Origin of the Hot Early Universe},
\npb{862}{2012}{587} [\arxiv{1202.6679}].




\bibitem{asfhi} G. Lazarides and C.~Pallis,
\textit{Probing the Supersymmetry-Mass Scale With F-term Hybrid
Inflation}, {\sl Phys.\ Rev.\ D {\bf 108}, no. 9, 095055 (2023)}
[\texttt{arXiv:2309.04848}].


\bibitem{susyr} C.~Pallis, {\it Gravity-mediated SUSY breaking, R symmetry,
and hyperbolic K\"ahler geometry,} {\sl Phys.\ Rev.\ D }{\bf 100},
no.~5, 055013 (2019) [\arxiv{1812.10284}]; C.~Pallis, {\it
SUSY-breaking scenarios with a mildly violated $R$ symmetry,} {\sl
Eur. Phys. J. C} \textbf{81}, no.~9, 804 (2021)
[\arxiv{2007.06012}].




\bibitem{nshi} L.~Wu, S.~Hu and T.~Li, {\it No-Scale $\mu$-Term Hybrid
Inflation,} {\sl Eur. Phys. J. C} \textbf{77}, no.~3, 168 (2017)
[\arxiv{1605.00735}].

\bibitem{buch1} W. Buchm\"uller, L. Covi and D. Delepine, {\it Inflation and supersymmetry breaking}, {\sl Phys. Lett. B }{\bf 491}, 183
(2000) [\hepph{0006168}].

\bibitem{stefan} S. Antusch, M. Bastero-Gil, K. Dutta, S.F. King and P.M.
Kostka, {\it Solving the eta-Problem in Hybrid Inflation with
Heisenberg Symmetry and Stabilized Modulus}, \jcap{01}{2009}{040}
[\arxiv{0808.2425}].

\bibitem{high} T. Higaki, K.S. Jeong and F. Takahashi, {\it Hybrid
inflation in high-scale supersymmetry} \jhep{12}{2012}{111}
[\arxiv{1211.0994}].

\bibitem{davis} P.~Brax, C.~van de Bruck, A.C.~Davis and S.C.~Davis,
{\it Coupling hybrid inflation to moduli} \jcap{09}{2006}{012}
[\hepth{0606140}]; S.C. Davis and M. Postma, {\it Successfully
combining SUGRA hybrid inflation and moduli stabilisation},
\jcap{04}{2008}{022} [\arxiv{0801.2116}]; S. Mooij and M. Postma,
{\it Hybrid inflation with moduli stabilization and low scale
supersymmetry breaking}, \jcap{06}{2010}{012} [\arxiv{1001.0664}].



\bibitem{mubaer}  K.J.~Bae, H.~Baer, V.~Barger and D.~Sengupta, {\it Revisiting the
SUSY $\mu$ problem and its solutions in the LHC era,} {\sl Phys.\
Rev.\ D }{\bf 99}, no.~11, 115027 (2019) [\arxiv{1902.10748}].


\bibitem{masiero} G.F. Giudice and A. Masiero, {\it A Natural Solution to
the mu Problem in Supergravity Theories}, {\sl Phys. Lett. B} {\bf 206}, 480 (1988).

\bibitem{soft} A. Brignole, L.E. Ib\'a\~nez and C. Mu\~noz, {\it Soft supersymmetry
breaking terms from supergravity and superstring models}, {\sl
Adv.\ Ser.\ Direct.\ High Energy Phys. }{\bf 18}, 125 (1998)
[\hepph{9707209}].


\bibitem{muhi} N.~Okada and Q.~Shafi, {\it $\mu$-term hybrid inflation and
split supersymmetry,} {\sl Phys. Lett. B} \textbf{775}, 348 (2017)
[\arxiv{1506.01410}]; M.U.~Rehman, Q.~Shafi and F.K.Vardag, {\it
$\mu$-Hybrid Inflation with Low Reheat Temperature and Observable
Gravity Waves}, {\sl Phys. Rev. D} \textbf{96}, no. 6, 063527
(2017) [\arxiv{1705.03693}].



\bibitem{dvali} G.R. Dvali, G. Lazarides and Q. Shafi, {\it Mu problem and
hybrid inflation in supersymmetric $SU(2)_{\rm L} \times
SU(2)_{\rm R}\times U(1)_{B-L}$,} {\sl Phys. Lett. B} {\bf 424},
{259} ({1998}) [\hepph{9710314}].





\bibitem{moduli} G. Kane, K. Sinha and S. Watson, {\it Cosmological Moduli and the
Post-Inflationary Universe: A Critical Review}, {\sl Int. J. Mod.
Phys. D } {\bf 24}, no.~08, 1530022 (2015) [\arxiv{1502.07746}].



\bibitem{baerh} K.J.~Bae, H.~Baer, V.~Barger and R.W.~Deal,
{\it The cosmological moduli problem and naturalness},
\jhep{02}{2022}{138} [\arxiv{2201.06633}].


\bibitem{full} M.~Endo, F.~Takahashi, and T.T.~Yanagida, {\it Inflaton Decay in
Supergravity,} {\sl Phys. Rev. D }\textbf{76}, 083509 (2007)
[\arxiv{0706.0986}].

\bibitem{nsrh} J. Ellis, M. Garcia, D. Nanopoulos and K. Olive, {\it Phenomenological
Aspects of No-Scale Inflation Models}, {\sl J. Cosmol. Astropart.
Phys. }{\bf 10}, 003 (2015) [\arxiv{1503.08867}].


\bibitem{antrh} Y.~Aldabergenov, I.~Antoniadis, A.~Chatrabhuti and H.~Isono,
{\it Reheating after inflation by supersymmetry breaking,} {\sl
Eur. Phys. J. C }\textbf{81}, no.~12, 1078 (2021)
[\arxiv{2110.01347}].



\bibitem{nsref} T.~Hasegawa \etal, {\it MeV-scale reheating temperature and
thermalization of oscillating neutrinos by radiative and hadronic
decays of massive particles,} \jcap{12}{2019}{012}
[\arxiv{1908.10189}].




\bibitem{wells} Y. Cui, M. Lewicki, D.E. Morrissey and J.D. Wells, {\it Probing
the pre-BBN universe with gravitational waves from cosmic
strings}, \jhep{01}{2019}{081} [\arxiv{1808.08968}];
Y.~Gouttenoire, G.~Servant and P.~Simakachorn, {\it Beyond the
Standard Models with Cosmic Strings} \jcap{07}{2020}{032}
[\arxiv{1912.02569}]; C.F.~Chang and Y.~Cui, {\it Gravitational
waves from global cosmic strings and cosmic archaeology,}
\jhep{03}{2022}{114} [\arxiv{2106.09746}].



\bibitem{pillado} P. Auclair \etal, {\it Probing the gravitational wave
background from cosmic strings with LISA,} \jcap{04}{2020}{034}
[\arxiv{1909.00819}].


\bibitem{ligo} R.~Abbott \textit{et al.} [LIGO Scientific, Virgo and KAGRA Collaboration],
{\it Constraints on Cosmic Strings Using Data from the Third
Advanced LIGO-Virgo Observing Run}, {\sl Phys. Rev. Lett. }{\bf
126}, no.~24, 241102 (2021) [\arxiv{2101.12248}].

\bibitem{wellspev} J.D.~Wells, {\it PeV-Scale Supersymmetry,} {\sl Phys. Rev. D}
\textbf{71}, 015013 (2005) [\hepph{0411041}].

\bibitem{strumia} E.~Bagnaschi, G.F.~Giudice,
P.~Slavich and A.~Strumia, {\it Higgs Mass and Unnatural
Supersymmetry,} \jhep{09}{2014}{092} [\arxiv{1407.4081}].

\bibitem{ant1} R. Kallosh and A. Linde, {\it Planck, LHC and
$\alpha$-attractors,} {\sl Phys. Rev. D} {\bf 91}, 083528 (2015)
[\arxiv{1502.07733}]; M.C.~Rom\~ao and S.F.~King, {\it
Starobinsky-like inflation in no-scale supergravity Wess-Zumino
model with Polonyi term} \jhep{07}{2017}{033}
[\arxiv{1703.08333}]; K. Harigaya and K. Schmitz, {\it Inflation
from High-Scale Supersymmetry Breaking,} {\sl Phys. Lett. B} {\bf
773}, 320 (2017) [\arxiv{1707.03646}]; I.~Antoniadis,
A.~Chatrabhuti, H.~Isono and R.~Knoops, {\it Inflation from
Supersymmetry Breaking,} {\sl Eur. Phys. J. C }\textbf{77},
no.~11, 724 (2017) [\arxiv{1706.04133}]; E.~Dudas, T.~Gherghetta,
Y.~Mambrini and K.A.~Olive, {\it Inflation and High-Scale
Supersymmetry with an EeV Gravitino}, {\sl Phys.\ Rev.\ D } {\bf
96}, no. 11, 115032 (2017) [\arxiv{1710.07341}]; Y.~Aldabergenov,
A.~Chatrabhuti and S.V.~Ketov, {\it Generalized dilaton-axion
models of inflation, de Sitter vacua and spontaneous SUSY breaking
in supergravity} {\sl Eur.\ Phys.\ J.\ C }{\bf 79}, no.~8, 713
(2019) [\arxiv{1907.10373}]; Y.~Aldabergenov, A.~Chatrabhuti and
H.~Isono, {\it $\alpha$-attractors from supersymmetry breaking,}
{\sl Eur. Phys. J. C }\textbf{81}, no.~2, 166 (2021)
[\arxiv{2009.02203}]; C.~Pallis, {\it Inflection-point sgoldstino
inflation in no-scale supergravity,} {\sl Phys. Lett. B}
\textbf{843}, 138018 (2023) [\arxiv{2302.12214}].

\bibitem{tamv} S.~Abel, A.~Dedes and K.~Tamvakis,
{\it Naturally small Dirac neutrino masses in supergravity,} {\sl
Phys. Rev. D }\textbf{71}, 033003 (2005) [\hepph{0402287}].


\bibitem{koichi} M. Endo \etal, {\it Moduli-induced gravitino
problem}, {\sl Phys. Rev. Lett.} {\bf 96}, 211301 (2006) [\hepph{
0602061}]; S. Nakamura and M. Yamaguchi, {\it Gravitino production
from heavy moduli decay and cosmological moduli problem revived},
{\sl Phys. Lett. B} {\bf 638}, 389 (2006) [\hepph{0602081}].




\bibitem{kolb} G.F. Giudice, E.W. Kolb and A. Riotto, {\it
Largest temperature of the radiation era and its cosmological
implications}, \prd{64}{2001}{023508} [\hepph{0005123}].


\bibitem{rh} C.~Pallis, {\it Massive particle decay and cold dark matter abundance},
{\sl Astropart. Phys.} \textbf{21}, 689 (2004) [\hepph{0402033}];
C. Pallis, {\it Cold Dark Matter in non-Standard Cosmologies,
PAMELA, ATIC and Fermi LAT}, \npb{751}{2006}{129}
[\hepph{0510234}].




\bibitem{micro} G. B\'{e}langer {\it et al.}, {\it MicrOMEGAs: A
Program for calculating the relic density in the MSSM}, \cpc{149}{2002}{103}
[\hepph{0112278}]; P. Gondolo {\it et al.}, {\it DarkSUSY:
Computing supersymmetric dark matter properties numerically},
\jcap{07}{2004}{008} [\astroph{0406204}].

\bibitem{qcd} C.~Pallis,
{\it Quintessential kination and cold dark matter abundance,}
\jcap{10}{2005}{015} [\hepph{0503080}].



\bibitem{plcp} N.~Aghanim {\it et al.} [\plk\ Collaboration],
{\it  Planck 2018 results. VI. Cosmological parameters}, {\sl
Astron. Astrophys. }\textbf{641}, A6 (2020); Astron.Astrophys. 652
(2021) C4 (erratum) [\arxiv{1807.06209}].

\bibitem{gws} M. Tristram \etal, {\it Improved limits on the tensor-to-scalar
ratio using BICEP and Planck}, {\sl Phys. Rev. Lett. }{\bf 127},
151301 (2021) [\arxiv{2112.07961}].

\bibitem{plcs0} P.A.R. Ade \etal\ [\plk\ Collaboration], {\it Planck 2015 results. XIII. Cosmological
parameters} {\sl Astron. Astrophys. }{\bf 594}, A13 (2016)
[\arxiv{1502.01589}].

\bibitem{bbngws} C. Caprini and D.G. Figueroa, {\it Cosmological Backgrounds of Gravitational Waves},
{\sl Class. Quant. Grav. }{\bf 35}, no.~16, 163001 (2018)
[\arxiv{1801.04268}].

\bibitem{ska}  G. Janssen \etal, {\it Gravitational wave astronomy with
the SKA,} {\sl PoS} {\bf AASKA14}, 037(2015) [\arxiv{1501.00127}].

\bibitem{thia} C. Boehm \etal\ [Theia Collaboration], {\it Theia: Faint
objects in motion or the new astrometry frontier,}
\arxiv{1707.01348}.

\bibitem{mares}  A. Sesana \etal, {\it Unveiling the gravitational universe
at $\mu-Hz$ frequencies,} {\sl Exper. Astron. }{\bf 51} 1333,
no.~3, (2021) [\arxiv{1908.11391}].

\bibitem{lisa}  P. Amaro-Seoane \etal\ [LISA Collaboration], {\it Laser
Interferometer Space Antenna,} \arxiv{1702.00786}.

\bibitem{tj}  W.-H. Ruan, Z.-K. Guo, R.-G. Cai and Y.-Z. Zhang, {\it Taiji
program: Gravitational-wave sources,} {\sl Int. J. Mod. Phys. A
}{\bf 35}, 2050075, no.~17 (2020)  [arxiv{1807.09495}].

\bibitem{tq}  J. Luo \etal\ [TianQin Collaboration], {\it TianQin: a
space-borne gravitational wave detector,} {\sl Class. Quant. Grav.
}{\bf 33}, 035010, no.~3, (2016) [\arxiv{1512.02076}].

\bibitem{bbo}  V. Corbin and N. J. Cornish, {\it Detecting the cosmic
gravitational wave background with the big bang observer,} {\sl
Class. Quant. Grav. }{\bf 23}, 2435 (2006) [{\tt\ftn
gr-qc/0512039}].

\bibitem{decig}  N. Seto, S. Kawamura and T. Nakamura, {\it Possibility
of direct measurement of the acceleration of the universe using
0.1-Hz band laser interferometer gravitational wave antenna in
space,} {\sl Phys. Rev. Lett. }{\bf 87}, 221103 (2001)
[\astroph{0108011}].

\bibitem{et}  B. Sathyaprakash \etal, {\it Scientific Objectives of
Einstein Telescope,} {\sl Class. Quant. Grav. }{\bf 29} 124013
(2012) [\arxiv{1206.0331}]; Erratum: {\sl Class. Quant. Grav.
}{\bf 30}, 079501 (2013)].

\bibitem{ce}  B. P. Abbott \etal\ [LIGO Scientific Collaboration],
{\it Exploring the Sensitivity of Next Generation Gravitational
Wave Detectors,} {\sl Class. Quant. Grav. }{\bf 34} no.~4, 044001
(2017) [\arxiv{1607.08697}].


\bibitem{allahbau} R.~Allahverdi, B.~Dutta and K.~Sinha, {\it Baryogenesis and
Late-Decaying Moduli,} {\sl Phys. Rev. D }\textbf{82}, 035004
(2010) [\arxiv{1005.2804}].

\bibitem{bau} M.M.~Flores, A.~Kusenko, L.~Pearce, and G.~White,
{\it Fireball baryogenesis from early structure formation due to
Yukawa forces,} {\sl Phys. Rev. D }\textbf{108}, no.9, 9 (2023)
[\arxiv{2208.09789}].


\bibitem{cbau} J. Garcia-Bellido \etal, {\it Nonequilibrium electroweak
baryogenesis from preheating after inflation}, {\sl Phys. Rev. D
60}, 123504 (1999) [{\ftn\tt hep-ph/9902449}]; L.M. Krauss and M.
Trodden, {\it Baryogenesis below the electroweak scale}, {\sl
Phys. Rev. Lett. }{\bf 83}, 1502 (1999) [\hepph{9902420}].

\bibitem{kolbdm} D.J.H.~Chung, E.W.~Kolb and A.~Riotto, {\it Nonthermal supermassive dark matter},
{\sl Phys. Rev. Lett. }\textbf{81}, 4048 (1998) [\hepph{9805473}];
D.J.H.~Chung, E.W.~Kolb and A.~Riotto, {\it Superheavy dark
matter,} {\sl Phys. Rev. D} \textbf{59}, 023501 (1998)
[\hepph{9802238}].

\bibitem{Raxiondm} H.S. Goh and M. Ibe, {\it R-axion detection at LHC}, {\sl J. High Energy Phys. }{\bf 03}, 049 (2009)
[\arxiv{0810.5773}]; Y. Hamada \etal, {\it More on cosmological
constraints on spontaneous R-symmetry breaking models}, {\sl J.
Cosmol. Astropart. Phys. }{\bf 01}, 024 (2014)
[\arxiv{1310.0118}].


\end{thebibliography}
\end{document}